\documentclass[fleqn,usenatbib]{mnras}
\usepackage{newtxtext,newtxmath}

\DeclareRobustCommand{\VAN}[3]{#2}
\let\VANthebibliography\thebibliography
\def\thebibliography{\DeclareRobustCommand{\VAN}[3]{##3}\VANthebibliography}
\defcitealias{2024MNRAS.527.5265C}{Paper I}

\usepackage{graphicx}
\usepackage{mathtools}
\usepackage{amsmath}
\usepackage[dvipsnames]{xcolor}
\usepackage{tabularx}
\usepackage{multirow}
\usepackage{float}

\usepackage{subcaption}

\newcommand\Msun{{\rm M}_\odot}
\newcommand\chimp{h^{-1}{\rm Mpc}}
\newcommand\chikp{h^{-1}{\rm kpc}}
\newcommand\Mstar{M_*}

\newcommand\hiimsun{h^{-2}{\rm M}_\odot}
\newcommand\himsun{h^{-1}{\rm M}_\odot}

\newcommand\zbest{z_{\rm best}}

\newcommand\zs{z_{\rm s}}
\newcommand\zl{z_{\rm l}}

\newcommand\Mstrue{M_{*, \rm t}}
\newcommand\Msobs{M_{*, \rm o}}

\newcommand\fsat{f_{\rm sat}}

\newcommand\logMstarlimit{\log\left[ M_{*, \rm limit}/ (h^{-2}\Msun) \right]}
\newcommand\logMstarh{\log\left[ M_{*} / (h^{-2}\Msun) \right]}

\newcommand{\avg}[1]{{\langle #1 \rangle}}
\newcommand{\ncen}{N_{\rm c}}
\newcommand{\nsat}{N_{\rm s}}
\newcommand{\Mmin}{M_{\rm min}}

\newcommand{\avMcen}{\avg{M_{\rm cen}}}
\newcommand{\logMmin}{\log M_{\rm min}}
\newcommand{\sigmalogM}{\sigma_{\log M}}
\newcommand{\rmd}{{\rm d}}
\newcommand{\kmspmpc}{{\rm kms^{-1}Mpc^{-1}}}

\newcommand{\photoz}{{\rm photo}-$z$\,}
\newcommand{\zt}{z_{\rm t}}
\newcommand{\zo}{z_{\rm o}}

\newcolumntype{C}{>{\centering\arraybackslash}p{1.4cm}}
\newcommand{\rmo}{{\rm o}}
\newcommand{\rmt}{{\rm t}}

\newcommand{\rmcrit}{{\rm crit}}
\newcommand{\pzmock}{Pz-mock~}
\newcommand{\pzmocks}{Pz-mocks}
\newcommand{\lspair}{${\rm l-s}$}

\title[Galaxy-dark matter connection in HSC]{Galaxy-dark matter connection from weak lensing in imaging surveys: Impact of photometric redshift errors}

\author[Chaurasiya et al.]{
Navin Chaurasiya,$^{1}$\thanks{E-mail: navin@iucaa.in }
Surhud More,$^{1,2}$\thanks{E-mail: surhud@iucaa.in}
Daichi Kashino,$^{3}$
Shogo Masaki,$^{4,5}$
Shogo Ishikawa$^{6}$
\\
$^{1}$Inter-University Centre for Astronomy and Astrophysics, Ganeshkhind, Pune 411007, IN\\
$^{2}$Kavli Institute for the Physics and Mathematics of the Universe (WPI), University of Tokyo, 5-1-5, Kashiwanoha, 2778583, JP\\
$^{3}$Division of Science, National Astronomical Observatory of Japan, 2-21-1 Osawa, Mitaka, Tokyo 181-8588, Japan\\
$^4$Department of Information Engineering and Institute for Advanced Studies in Artificial Intelligence, Chukyo University, Toyota, Aichi 470-0348, Japan\\
$^5$Department of Physics, Nagoya University, Nagoya, Aichi 464-8601, Japan\\
$^{6}$Department of Liberal Arts and Basic Sciences, College of Industrial
Technology, Nihon University, Narashino, Chiba 275-8576, Japan}

\date{Accepted XXX. Received YYY; in original form ZZZ}

\pubyear{2023}

\begin{document}
\label{firstpage}
\pagerange{\pageref{firstpage}--\pageref{lastpage}}
\maketitle

\begin{abstract}
The uncertainties in photometric redshifts and stellar masses from imaging surveys affect galaxy sample selection, their abundance measurements, as well as the measured weak lensing signals. We develop a framework to assess the systematic effects arising from the use of redshifts and stellar masses derived from photometric data, and explore their impact on the inferred galaxy-dark matter connection. We use galaxy catalogues from the UniverseMachine (UM) galaxy formation model to create \pzmock galaxy samples that approximately follow the redshift errors in the Subaru HSC survey. We focus on galaxy stellar-mass thresholds ranging from $\logMstarh$ from $8.6$ to $11.2$ in steps of 0.2 dex within two redshift bins $0.30-0.55$ and $0.55-0.80$. A comparison of the \pzmock samples to true galaxy samples in UM shows a relatively mild sample contamination for thresholds with $\logMstarlimit<10.6$, while an increasing contamination towards the more massive end. We show how such contamination affects the measured abundance and the lensing signal. A joint HOD modelling of the observables from the \pzmock compared to the truth in the UM informs the systematic biases on the average halo masses of central galaxies in the HSC survey. Even with a reasonably conservative choice of \photoz{} errors in \pzmocks{}, we show that the inferred halo masses deduced from the HSC galaxies for low-mass thresholds will have a systematic bias smaller than 0.05 dex. Beyond $\logMstarlimit=10.6$, the inferred halo masses show an increasing systematic bias with stellar mass, reaching values of order $0.2$ dex, larger than the statistical error.
\end{abstract}


\section{Introduction} \label{intro}

In the concordance cosmological model, dark energy and dark matter
dominate the energy density of the Universe, and their interplay
determines the rate of growth of structure in the Universe. The
parameters that describe the cosmological model have been constrained
with unprecedented precision via a wealth of observations, such as the
anisotropies in the cosmic microwave background \citep{2020A&A...641A...6P}, geometric probes such
supernovae \citep{2018ApJ...859..101S, 2024arXiv240102929D} and baryon acoustic oscillations \citep{2021PhRvD.103h3533A, 2022PhRvD.105d3512A, 2023NatAs...7.1259X, 2024arXiv240403002D}, as well as probes such as
abundance of galaxy clusters \citep{2023MNRAS.522.1601C}, the clustering of galaxies and their galaxy-galaxy lensing signal \citep{10.1093/mnras/sts525, 2017MNRAS.465.4204C, 2022MNRAS.509.3119Z, 2022PhRvD.105b3520A, 2023PhRvD.108l3517M, 2023A&A...675A.189D} and the cosmic shear \citep{2022PhRvD.105b3514A, 2023PhRvD.108l3518L, 2023PhRvD.108l3519D} which are sensitive
to the growth of large-scale structures. Large imaging surveys, such as
the Dark Energy Survey \citep[DES;][]{2018ApJS..239...18A}, Kilo-Degree
Survey \citep[KiDS;][]{2015A&A...582A..62D}, and Hyper Suprime-Cam
Subaru Strategic Program \citep[HSC-SSP;][]{2018PASJ...70S...8A} and
spectroscopic surveys, such as the Sloan Digital Sky Survey \citep[SDSS;][]{2000AJ....120.1579Y} and the
Dark Energy Spectroscopic Instrument \citep[DESI;][]{2023arXiv230606308D} have led to considerable
statistically large samples which have improved the accuracy in the
measurement of large-scale structure \citep{2020MNRAS.495.3695K, 2023PhRvD.108l3520M}.

The statistical errors can be further improved with the use of
information from the quasi-linear or non-linear scales
\citep{2014MNRAS.444..476R, 2019MNRAS.484..989W, 2019MNRAS.488.5771L,
2022MNRAS.509.1779L, 2022MNRAS.510.5376S, 2022PhRvD.106j3530P,
2022MNRAS.515.2612M, 2022arXiv221100723H, 2023A&A...675A.189D, 2023MNRAS.520.5373L}. The use of these scales
often necessitates an understanding of the connection between the
baryonic components of galaxies and their dark matter halos
\citep{2002ApJ...575..587B, PhysRevD.86.083540, 2014PhRvD..90l3522S}.
Both galaxy clustering \citep{2011ApJ...736...59Z, 2016MNRAS.460.1457S, 2016MNRAS.460.1173R} and galaxy-galaxy lensing \citep{2022A&A...668A..12L, 2024MNRAS.527.5265C} on small scales have
been used separately as well as together to constrain the galaxy-dark
matter connection and its evolution with redshift \citep{2011ApJ...738...45L, 2012ApJ...744..159L, 2015MNRAS.449.1352C, 2015ApJ...806....2M, 10.1093/mnras/stac472, 2016MNRAS.457.4360Z}

Deep imaging surveys such as the Subaru HSC survey, allow galaxies
with stellar masses ($\Mstar$) as low as $10^9 \hiimsun$ to be detected even at
redshifts as high as $z=1$. \citet[][I20 hereafter]{2020ApJ...904..128I} measured the clustering of HSC galaxies in
stellar mass threshold samples and fitted an analytical halo model
framework to the data. Their analysis allowed them to infer the galaxy-dark matter connection and its evolution over the redshift range
$[0.3, 0.8]$. In a follow-up study, \citet[see][hereafter, Paper
I]{2024MNRAS.527.5265C} measured the weak gravitational lensing
signals (WLSs) due to these galaxies acting as lenses for the background source galaxies from the Subaru HSC survey, to directly probe the dark matter halo masses of the lensing galaxies and thus improving the potential statistical power further.

Harnessing the increased statistical power, however, requires a deeper
understanding of various systematic biases, to avoid improving the
statistical precision at the cost of loss in accuracy. One such
systematic uncertainty is related to the errors in the redshift
estimates of galaxies as determined from multi-wavelength fluxes from
imaging surveys.

Even a simple statistical error in the individual photometric
redshifts (photo-$z$s) can result in overall biases in the measured
distributions of various correlated galaxy properties in a non-trivial
manner, such as luminosity or the stellar mass function (SMF), as well
as the integrated number densities of galaxies in threshold bins,
which have been traditionally used in halo model studies in addition
to the clustering and lensing measurements.   

The photo-$z$ errors can also propagate to distinct observational
probes in different manners. For example, \citet{2021MNRAS.501.3309Z}
discuss the impact on clustering of photometric luminous red
galaxies (LRGs) in DESI, \citet{2018MNRAS.477.3892C} on clustering and
baryon acoustic oscillation, and \citet{2019MNRAS.486.2730A} on
lensing peak counts and power spectrum and their role in constraining
cosmological parameters. In galaxy-galaxy lensing analysis, the
tangential shear of galaxies measured as a function of projected
distances for photometric lens galaxies depends on individual lens
redshift information to produce accurate stacking of the weak lensing
signal across the radial bins. Uncertainties in lens redshifts cause
mixing of lens-source (\lspair) pairs across multiple radial bins and also assign
incorrect scaling to the signal contribution from each lens
\citep[see][section 8]{2024MNRAS.527.5265C}.  
Moreover, the uncertainties in galaxy properties, which also lead to the problem of sample selection bias and incompleteness in photometric datasets, pose a serious challenge to bridging
the gap between empirical observations and theoretical predictions of
both galaxy evolution and cosmology.

In this work, we examine the extent to which the abundance and the
galaxy-galaxy lensing signals of samples of galaxies defined based on
thresholds in observed stellar mass can be affected due to errors in
photometric redshifts. We will model these systematically biased observables in
the standard HOD model framework and estimate the potential level of
bias in the inference of the galaxy-dark matter connection and its
evolution. For this purpose, we will construct mock galaxy samples
from a leading model of galaxy formation, the {\sc UniverseMachine}, and
include errors in redshift as well as correlated errors in stellar masses to similar ballpark levels as expected from the photometric
catalogue {\sc Mizuki} of the HSC survey.

This paper is organised as follows: We describe the simulation data in
Section~\ref{data} and our methodology of the estimation of photo-$z$
uncertainties in the HSC survey and the construction of resultant mock
stellar mass threshold sample in Section~\ref{methodology}. We
describe the contamination in the mock samples in
Section~\ref{lens_sampling} and the formalism to derive the WLS for
these \pzmocks{} and underlying true samples in
Section~\ref{measurement}. Our theoretical modelling framework is
described in Section~\ref{theoretical_modelling} and the results of
measurements and model fitting are given in Section~\ref{results}. We
describe the caveats and challenges in the current work and its future
outlook in Section~\ref{challenges} and summarise the paper and
conclude in Section~\ref{summary}.

\section{Data} \label{data}
\subsection{Simulation}
For the purpose of carrying out a realistic emulation of the photometric redshift errors that are present in the HSC data, we start from galaxy catalogues derived from the {\sc UniverseMachine} \citep[UM;][henceforth B19]{2019MNRAS.488.3143B} Data Release 1 (DR1), a galaxy formation model built on top of halo merger trees from a collisionless simulation. The particular galaxy catalogues that we utilise in this work are based on \textit{Bolshoi–Planck} \citep{2016MNRAS.457.4340K, 2016MNRAS.462..893R}, which corresponds to a box size of $250 \chimp$, simulated with $2048^3$ particles in a flat $\Lambda$CDM cosmology with parameters, matter density parameter $\Omega_{\rm m}=0.307$, baryon density parameter $\Omega_{\rm b}=0.048$, the power law slope of the initial power spectrum $n_{\rm s}=0.96$, and its amplitude characterised by the parameter $\sigma_8=0.823$, and $h=0.678$, where $h=H_0/(100 \kmspmpc)$. 

The galaxy formation model uses properties of dark matter halos and their assembly histories from the simulation based on the {\sc ROCKSTAR} halo finder \citep{2013ApJ...762..109B} and {\sc Consistent Trees} code \citep{2013ApJ...763...18B} for finding halos and construction of merger trees respectively. It assumes the \citet{2003PASP..115..763C} stellar IMF, the \citet{2003MNRAS.344.1000B} stellar population synthesis model, and the \citet{2000ApJ...533..682C} dust law to convert model stellar masses to luminosities in different wavelength bands to calibrate their galaxy formation model to various observations in the literature. These assumptions are compatible with the analyses of \citetalias{2024MNRAS.527.5265C} and HSC-{\sc Mizuki} \citep{2015ApJ...801...20T} photometric redshift catalogues used therein respectively. The UM catalogues include the positions and stellar masses of galaxies based on the best-fitting model from B19 constrained from various observational constraints discussed in section 2.2 of B19. 

We use the UM catalogue of the snapshot closest to the median redshift of the lowest galaxy stellar mass threshold $(\log M_{\rm *, limit})$ samples of each redshift bin, $z_1$ ($a=0.683373$) and $z_2$ ($a=0.597310$), defined in the Table~1 of \citetalias{2024MNRAS.527.5265C}. The median redshift is based on the \textit{best} estimate of the redshift ($\zbest$) of the individual galaxies given in the {\sc Mizuki} photometric catalogue of the HSC survey. We have checked that the small variation in the median redshift of different stellar mass threshold samples has an impact on the model observable which is much smaller than the statistical error, and hence we ignore it. While constraining the model, B19 also include systematic and statistical uncertainties in the observed values of stellar masses. However, such errors are expected to be dependent upon the survey specification and hence we use only the true stellar masses in our work so that we can assign observed estimates of the galaxy properties based on the characteristics of the HSC survey. 

The properties of galaxies, such as their stellar mass and the photometric redshift, are inferred from the observed fluxes in multiple bands typically by fitting a spectral energy distribution (SED). These physical properties and their errors depend upon the statistical errors in the photometric measurements across bands as well as systematic uncertainties arising out of degeneracies in the various modelling ingredients. Uncertainties in the modelling assumptions of the stellar population synthesis, metallicity, dust, star formation history as well as the stellar initial mass function can affect the templates for the SEDs. This results in degeneracies and correlated errors in the estimates of the physical properties of these galaxies. These uncertainties in inferred stellar masses can result in Eddington bias when galaxies are sampled to satisfy a threshold of observed stellar mass.

\section{Methodology} \label{methodology}
We develop a methodology to study the impact of uncertainties in the photometric redshifts of lenses on the galaxy-halo connection. In particular, we focus on the systematic biases imprinted on the measured WLS as well as the stellar mass-halo mass relation (SMHM) and its evolution as a function of redshift inferred from the simplistic modelling of the signal. Although we currently focus on the HSC survey, this methodology should be more generally applicable and could be used for the forward modelling of the WLSs in large photometric surveys. 

Our methodology can be summarised as follows. We estimate the uncertainty in the photometric redshifts of the lens galaxies that we use by comparing them with catalogues that either have spectroscopic redshifts or better-quality photometric redshifts. We imprint this estimated uncertainty on the redshifts of galaxies in the {\sc UniverseMachine} and propagate it to the inferred physical properties of galaxies such as their observed stellar mass. With such catalogues in hand that mimic the systematic uncertainties in the observations, we can now create mock samples of galaxies that are binned based on their observed redshift and stellar masses. We compute the WLS that will be measured around such galaxies taking into account their observed stellar masses, as well as their observed lens redshifts. We then infer the stellar mass-halo mass relation based on the HOD modelling of the WLS, which ignores any of the systematic uncertainties of the redshifts of the lens. Further, we compare this stellar mass-halo mass relation to the true relation in the galaxy catalogue to understand the extent of the systematic bias in the inference. We describe each of these steps in detail below.

\subsection{Accuracy of the photometric redshifts} \label{photoz_accuracy}
We estimate the accuracy of the photometric redshifts of the stellar mass threshold lens galaxies in HSC by utilising the 30-band photometric redshift information from the COSMOS field \citep{2016ApJS..224...24L}. The same galaxies were also observed by the HSC survey \citep[][]{2017arXiv170600566T} and their photometric redshifts \citep{2018PASJ...70S...9T, 2020arXiv200301511N} based on the HSC photometric data alone are also available publicly as part of Public data release 2 \citep{2019PASJ...71..114A}. Given the larger number of bands which sample the galaxy spectrum, we consider the COSMOS photometric redshifts to be the ground truth for the purpose of this paper. In reality, the COSMOS photometric redshifts for the relevant magnitude bins and at the median redshift of HSC ($\sim 0.8$) would have a scatter in $\Delta z = z_{\rm cosmo}-z_{\rm spec}$ of less than $0.05$ \citep{2016ApJS..224...24L} which are expected to be smaller compared to the photometric redshift errors in HSC, as these errors are expected to add in quadrature. 

\begin{figure}
    \centering
    \begin{subfigure}[b]{\columnwidth}
        \includegraphics[width=\columnwidth]{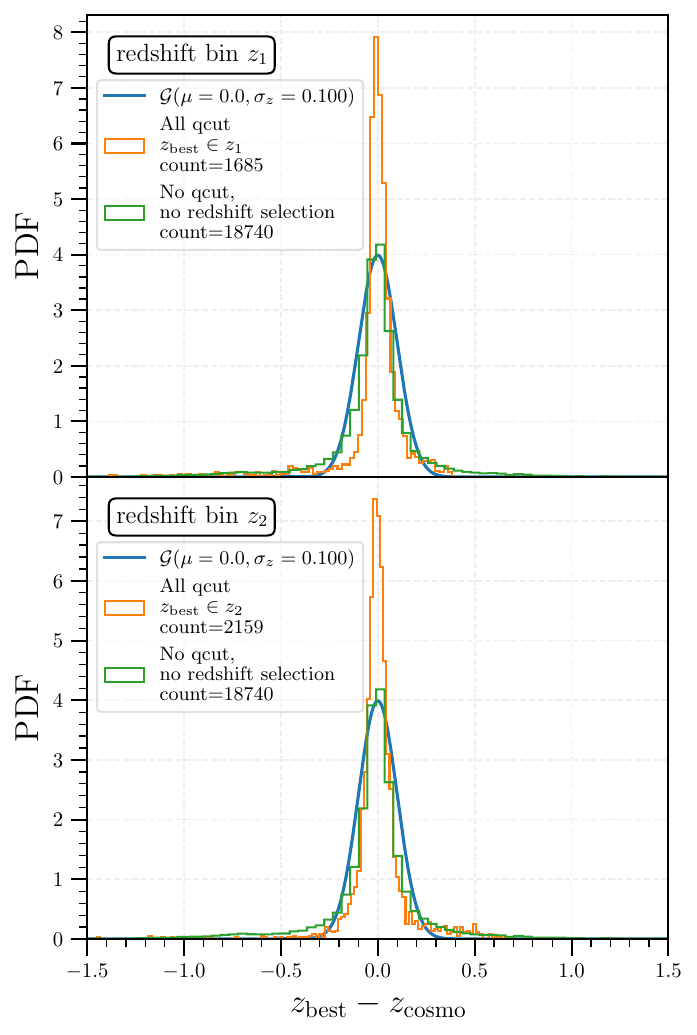}
    \end{subfigure}
    \caption{The orange histogram in the top and bottom panels show the PDF of the \photoz{} difference $(z_{\rm best}-z_{\rm cosmo})$ in the redshift bins 1 and 2 respectively for similar quality galaxies used as lenses in \citetalias{2024MNRAS.527.5265C}. In each panel, the green histogram shows the same PDF obtained from all the HSC galaxies in the COSMOS field without applying any lens quality cut or redshift bin selection. The blue curve is a Gaussian of width 0.1, shown to guide the eye around the green curve. 
    }
    \label{fig:hsc_photoz_cosmosz}
\end{figure}

We use the galaxies in HSC in the COSMOS area which have been cross-matched with galaxies in the COSMOS \photoz{} catalogues and provided by the HSC weak lensing team. These catalogues correspond to galaxies as faint as $i<24.5$, approximately similar magnitude cuts corresponding to our lens galaxy samples. The green histogram in both panels of Fig.~\ref{fig:hsc_photoz_cosmosz} shows the distribution of the difference between the HSC redshift and the COSMOS redshift, i.e., $\zbest-z_{\rm cosmo}$. The blue solid line in both panels shows a Gaussian with a width equal to $0.1$ shown to guide the eye. The HSC galaxies have a wider tail than the blue solid line.

However, in reality, we would like to characterise the galaxy samples in \citet[][henceforth I20]{2020ApJ...904..128I} which we used in \citetalias{2024MNRAS.527.5265C} for our lensing analysis, where several selection cuts were applied. I20 reject galaxies in the HSC catalogue that have been flagged for failure in the stellar mass estimation, or which do not satisfy the magnitude selection cuts applied (see sec 2.2 in I20) in any of the five HSC-wide bands ($g\leq26.0$, $r\leq25.6$, $i\leq25.4$, $z\leq24.2$, $y\leq23.4$ and $\ge 18.0$ in all bands)\footnote{Although these cuts seem to be deeper in the $i$-band than the COSMOS catalogue provided by the HSC weak lensing team, we have confirmed that after the stellar mass and redshift cuts, the galaxies that survive have $i<24.5$.}. The quality of the HSC photo-$z$s was also ensured by requiring {\sc photoz\, median} $\leq 3.0$, {\sc photoz\_risk\_best} $\leq 1$ and by removing the galaxies at the edges of the redshift within one standard deviation error (using {\sc photoz\_err68\_min} and {\sc photoz\_err68\_max} from {\sc Mizuki}) as described by the `trimming' procedure in Section 2.3.2 of I20. 

The resultant distribution of the difference in the redshift from HSC photometry to the redshift inferred from COSMOS 30-band photometric redshifts after the application of all these selection cuts for the two redshift bins $0.3 \leq \zbest \leq 0.55$ ($z_1$ bin) and $0.55 \leq \zbest \leq 0.80$ ($z_2$ bin) are shown in the two panels of Fig.~\ref{fig:hsc_photoz_cosmosz} with an orange histogram. The quality cuts ensure that the distributions are much narrower than that of all galaxies (shown in green).

We also observe that these distributions have significant kurtosis and to account for it, we will conservatively use an error of $\sigma_z=0.1$ which captures most of the wings of the green histogram although it misses the peak at the centre. 
Given our conservative choice for $\sigma_{\rm z}$, we expect the systematic effects we infer using this procedure to be a conservative estimate. We have also checked that these distributions do not change even after applying further cut on the minimum stellar mass threshold used in either redshift bin.

We note that in reality, the errors in the redshifts of galaxies are also expected to be correlated with the fluxes of the galaxies, and subsequently their stellar masses. However, due to a lack of number statistics owing to the limited COSMOS area in which we can conduct this exercise, we do not explore these dependencies in this paper and defer such investigations to the future when a larger sample may become available. If the differences are quantified in an even more nuanced manner, our methodology can be adapted in a straightforward manner.

\subsection{Mock catalogue with photometric redshift errors}
\label{subsec:mock_catalogue_with_photoz}
We use the galaxy catalogues from the {\sc UniverseMachine} for generating mock samples of galaxies including the photometric redshift errors derived above. We do not have galaxy catalogues on a light cone. Therefore, we span the required redshift range of interest by replicating the UM galaxy catalogues multiple times along the line-of-sight with periodic boundary conditions.\footnote{Note that we will not be actually measuring the lensing signal from the UM box, and thus repetition of the structure around galaxies along a given line-of-sight is not an issue.} For redshift bin 1 (2), we replicate the galaxy snapshot closest to the median redshift of the galaxy sample in the lowest threshold stellar mass bin, and cover a wider true redshift range from $0.02<z<1.1$ ($0.3<z<1.2$). In doing so, we have ignored the redshift evolution in the galaxy properties and their connection with the underlying dark matter halos in the redshift range of interest.

We work in the plane-parallel approximation. Each galaxy gets a true redshift, $\zt$, based on its comoving distance away from the observer. We sample the observed redshift for each galaxy, $\zo$ from a Gaussian distribution centred at $\zt$ with a $\sigma_z=0.1$. We have ignored peculiar velocities given the large magnitude of the photometric redshift uncertainty. We assign each galaxy a weight, $w_{\rm lc} \propto \chi^2$, where $\chi$ is the comoving distance along the line of sight from the observer. This weight allows us to mimic the light cone effect \footnote{The light cone effect is related to the fact that in a survey that spans a particular area on the sky, the comoving extent of the survey perpendicular to the line-of-sight will be smaller at smaller redshift.} while using the galaxies in the entire box in the plane parallel approximation. We will use this weight throughout the manuscript in all our figures concerning \pzmock samples constructed in Section~\ref{lens_sampling}.

The stellar masses of galaxies and the photometric redshifts have been inferred in the HSC public data release using the five-band $grizy$ data with the help of the template fitting code {\sc Mizuki} \citep{2015ApJ...801...20T}. Given that stellar mass estimates will have to rely on the estimated redshift to convert observed fluxes into a luminosity, and the observed colours to infer the SED, we expect the errors on the photometric redshifts to be highly correlated with the stellar mass. Therefore, we assign an observed stellar mass, $\Msobs$, for all mock galaxies such that
\begin{align}
    \Msobs = \Mstrue \left[\frac{d_{\rm L}(\zo)}{d_{\rm L}(\zt)}\right]^2\,,
    \label{eq:mstel_obs}
\end{align}
where $\Mstrue$ is the true stellar mass of the galaxy as given in the UM catalogue, i.e., the stellar mass assigned by the {\sc UniverseMachine}, and $d_{\rm L}(z)$ is the luminosity distance to a given redshift $z$. The use of this equation effectively guarantees that the flux of the observed galaxy is effectively preserved even though the observed redshift is different from the true redshift of the galaxy. We will comment on how a departure from this assumption will affect the results we show, especially with regard to sample selection, which we discuss next.

\begin{figure}
    \centering
    \includegraphics[width=\columnwidth]{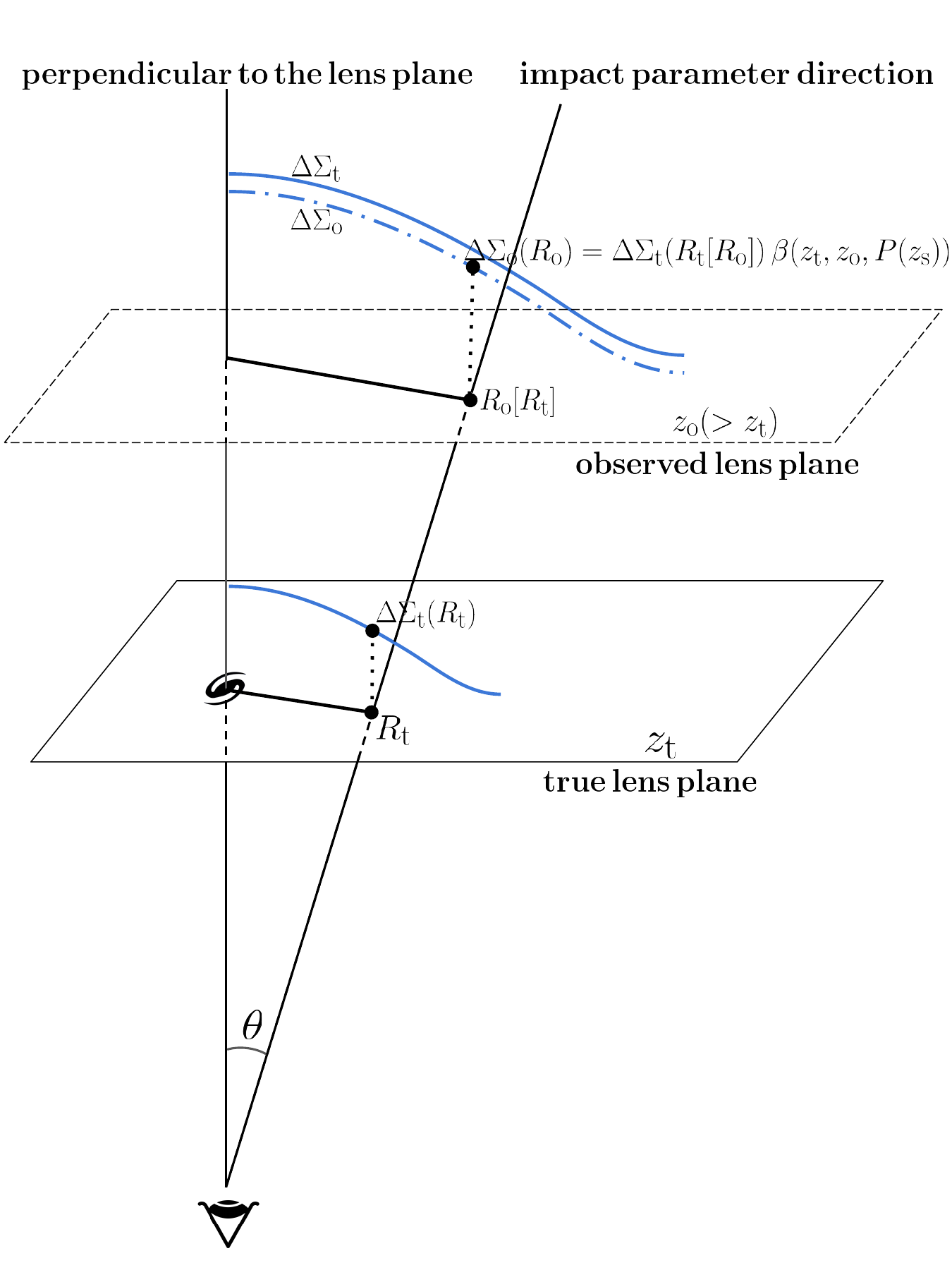}
    \caption{\textbf{Impact of the redshift error on the observed lensing signal $\Delta\Sigma_\rmo$ as a function of projected separation $R$: The case of a single lens galaxy with $\zo>\zt$}\\ 
    The eye at the bottom denotes the observer end connected by a straight line along the line-of-sight (LOS) to a lens galaxy situated at the true redshift $\zt$. 
    Assuming a thin lens approximation, we draw the true lens plane, a sky-projected plane drawn perpendicular to the line-of-sight to the galaxy at its redshift; and the observed lens plane, a similar projected plane imagined at the observed redshift $\zo$ of the galaxy along the extension of the LOS. 
    We show the WLS profile of a hypothetical lens galaxy as a function of projected distance $\Delta\Sigma (R)$ by a blue solid curve plotted in a plane defined by the cross product of lens plane normal and the projected radius. Also, the solid blue curve in the observed lens plane denotes the signal as if the same signal from $R_\rmt$ was assigned to $R_{\rmo}[R_\rmt]$. The dash-dot blue curve gives the signal that gets observed after accounting for \photoz{} errors. Note that the scaling factor $\beta$ can vary differently for different \lspair{} pairs. 
    }
    \label{fig:demo_figure}
\end{figure}

\section{Magnitude threshold samples of mock galaxies} \label{lens_sampling}

\begin{figure*}
    \centering
    \includegraphics[width=0.7\textwidth]{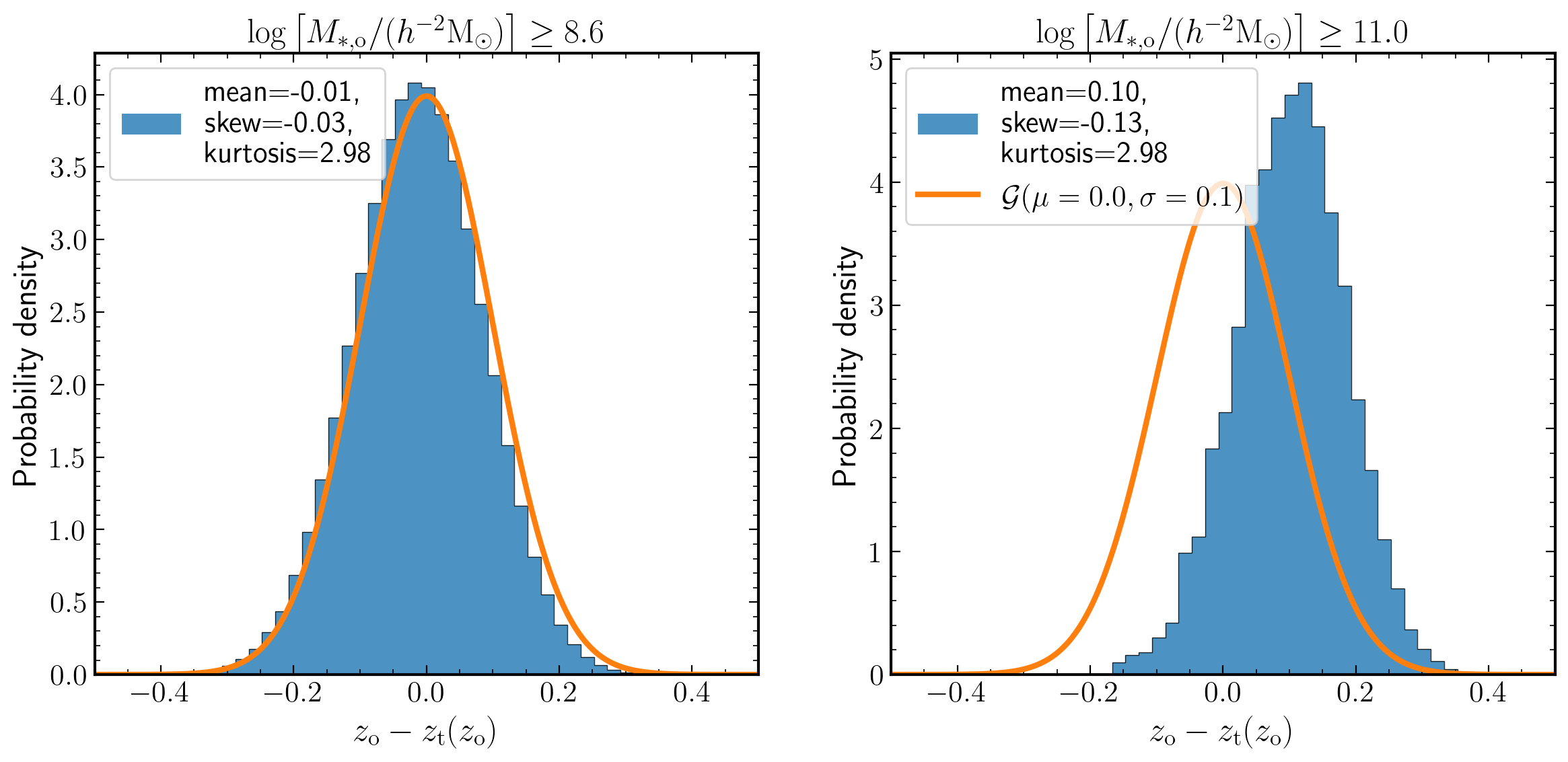}
    \caption{\pzmock samples in redshift bin $z_1$: The left and right panels correspond to the thresholds of 8.6 and 11.0 respectively as indicated in the title of each panel. The orange Gaussian curve represents the inputted redshift error profile in the UM galaxies. Deviations of the histogram from the orange curve quantify the level of redshift contamination in different threshold samples in the \pzmock. Refer to Section~\ref{lens_sampling} for discussion.}
    \label{fig:diff_z}
\end{figure*}

\begin{figure*}
    \centering
    \begin{subfigure}[b]{0.49\textwidth}
        \includegraphics[width=\columnwidth]{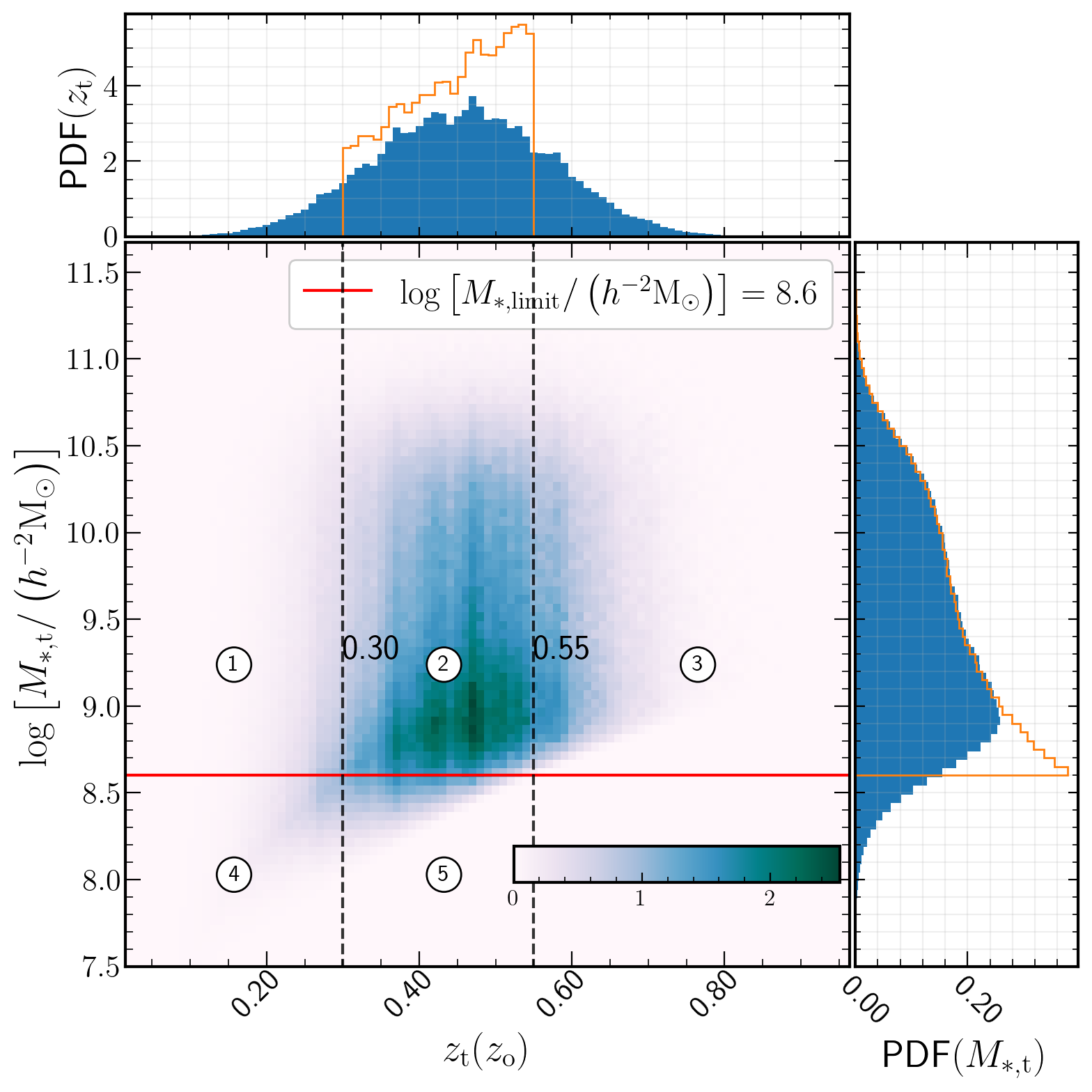}
        \caption{$z_{1}$: $\zo \in \left[0.30,0.55\right)$; $\log \left[ M_{*, \rm o}/(h^{-2} M_\odot ) \right] \geq 8.6$}
        \label{fig:smt_of_HSClike_8.6}
    \end{subfigure}
    \begin{subfigure}[b]{0.49\textwidth}
        \includegraphics[width=\columnwidth]{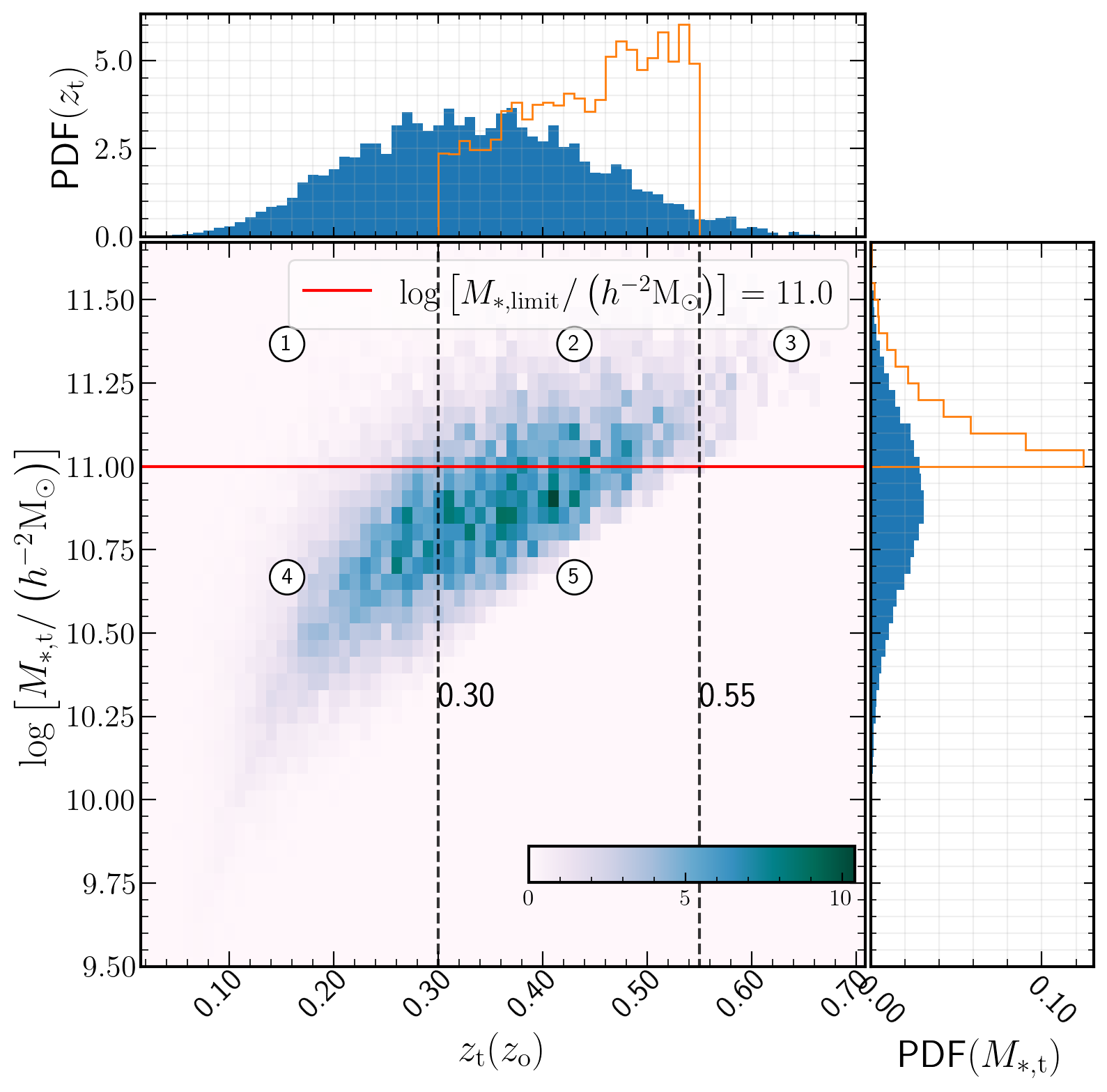}
        \caption{$z_{1}$: $\zo \in \left[0.30,0.55\right)$; $\log \left[ M_{*, \rm o}/(h^{-2} M_\odot ) \right] \geq 11.0$}
        \label{fig:smt_of_HSClike_11.0}
    \end{subfigure}
    \caption{Representation of the \pzmock samples in true lens plane: left and right panels correspond to thresholds of 8.6 and 11.0 respectively as indicated in the title of each panel. The threshold of the sample is shown by the red horizontal line while the vertical dotted lines are plotted to guide the human eyes for the supposed range of the redshift bin $z_1$ in which \pzmock data is sampled. The contamination of the redshift bin can be seen coming from either side of the dotted lines, and this causes a correlated contamination of the stellar mass threshold as well. Numbers within circles mark different contamination regions in the plot, refer to Section~\ref{lens_sampling} for discussion.}
    \label{fig:smt_of_HSClike_samples}
\end{figure*}

We construct \pzmock galaxy samples in identical thresholds of stellar masses defined in Table~1 of \citetalias{2024MNRAS.527.5265C} within the two redshift bins of our interest. For this purpose, we treat the mock redshifts and stellar mass estimates ($\zo$, $M_{*, \rm o}$) as equivalent to the ($\zbest$, $\Mstar$) parameters of the {\sc Mizuki} catalogue used in \citetalias{2024MNRAS.527.5265C}. In figures \ref{fig:diff_z}-\ref{fig:abundance}, we compare the true properties of these samples with the observed and also present the resultant contamination in the mock threshold samples. We expect that a similar level of contamination is likely to be present in the real HSC samples that were analysed in \citetalias{2024MNRAS.527.5265C}. The samples of galaxies based on the true redshift $\zt$ and thresholds of true stellar mass $\Mstrue$ that represent the underlying truth in the UM, will be denoted as UM-mock samples\footnote{Note that, we could also simply refer to these as UM samples, but we have maintained similarity in construction of \pzmock and UM-mock galaxy samples and the further computations to be performed on them.}. 

In Fig.~\ref{fig:diff_z}, we first show the PDF of the difference in the observed and true redshifts of galaxies residing in two different stellar mass threshold bins within the bin $z_1$. The two panels correspond to galaxies in the lowest and the highest stellar mass threshold bins and correspond to $\logMstarlimit$, of $8.6$ and $11.0$, respectively. The naive expectation would be that these distributions would match the input Gaussian with $\sigma_z=0.1$ that we assumed for the $P(\zo-\zt)$ distribution. However, as we are plotting the distribution for galaxies that enter a given threshold sample, rather than the full sample of galaxies in the entire mock, we expect that the asymmetric nature by which galaxies eventually end up in our sample will distort this distribution. For the lowest threshold bin, we see a relatively minor difference in this distribution from the input Gaussian distribution we assumed between $\zt$ and $\zo$. There is a slight preference for higher redshift galaxies entering our sample than lower redshift galaxies leaving the sample. However, as we move to the highest stellar mass threshold bin, this trend is reversed. In this bin, more low redshift galaxies enter the bin than higher redshift galaxies, thus shifting the mean of the difference in a positive direction.\footnote{The HSC samples of \citetalias{2024MNRAS.527.5265C} may not have \photoz{} errors of this magnitude but the overall sample contamination would have an additional component from errors in stellar masses independent of their \photoz{} errors. See discussion in Sections~\ref{subsec:mock_catalogue_with_photoz} and \ref{challenges}.}

These differences can be better understood by looking at the samples constructed using observed stellar mass threshold but in the true redshift true stellar mass plane, as shown in Fig.~\ref{fig:smt_of_HSClike_samples}.
The two panels correspond to the two threshold samples, with $\logMstarlimit$ equal to 8.6 and 11.0, respectively.
The larger subpanel shows the 2-D probability distribution of galaxies, while the blue-filled histograms in the top and right subpanels show the corresponding 1-D distributions of each of the samples in true redshift and true stellar mass, respectively. The orange colour histograms in these subpanels are the distributions expected for the true samples. The observed stellar mass threshold used to select each sample is marked by a horizontal red line and the limits of the redshift bin are shown as two vertical dotted lines. Given that the observed stellar masses are related to the true stellar masses of the galaxies by Eq.~\ref{eq:mstel_obs}, we expect galaxies from different regions to enter our observed sample.

These regions are marked from 1-5 in each of the main subpanels of Fig.~\ref{fig:smt_of_HSClike_samples}. Our region of interest is Region 2 which corresponds to galaxies in the correct true stellar mass threshold and the true redshift bin under consideration. Galaxies from the rest of the regions contaminate the sample. Galaxies in region 1 (3), have the correct true stellar masses that we want, but have a true lower (higher) redshift than the bin of interest and are scattered into the sample due to photometric redshift errors. On the other hand, galaxies in regions 4 and 5 have lower stellar masses than the threshold, but enter our sample as their observed stellar masses get boosted above the threshold due to errors in photometric redshifts that put them at higher redshifts.

Let us consider the interface between Regions 2 and 3. There are two main effects at play at this boundary which scatter galaxies in and out of our observed stellar mass threshold and redshift bin due to Eddington bias -- (i) the volume effect (more galaxies at true higher redshifts due to the light cone), which results in more galaxies that scatter into the sample due to assignment of lower redshifts, and  (ii) the stellar mass function effect (more galaxies at true smaller stellar masses), which causes more galaxies from the correct redshift bin to scatter out of the sample due to assignment of a larger photometric redshift. These effects counterbalance each other, and one of them is larger than the other, depending upon the sample under consideration. Similar competing effects are at play at the interfaces of the other regions. For the stellar mass bins with a lower threshold, a large portion of the sample is composed of galaxies that are in their correct redshift and stellar mass bin, but for the higher stellar mass threshold, the stellar mass function which is steeply declining, implies that most of the galaxies belong to lower stellar mass thresholds and scatter into the sample, especially from the interfaces of region 2 with regions 4 and 5, respectively. 

The shift in the 1-d distributions in stellar masses and redshifts as seen in the top and right subpanels by comparing the blue-filled histogram with the orange, show the combined effect of the photometric redshift errors on the samples. We see that there is quite a large difference in the distribution of the redshifts and stellar masses in these samples. However, we note that the distribution of the observed properties (redshift and stellar mass) of the stellar mass threshold samples compared to that of the true properties on which the samples are intended to be selected are very similar and thus the shape of the distribution itself is not a diagnostic of the extent of contamination of the sample.

Our assumption of a perfect correlation between the redshift errors and stellar mass errors will lead to a larger associated spread in the stellar mass direction than what might be present in the HSC catalogue. In the absence of such a correlation, we would expect the bottom envelope of the distribution of the true stellar mass and redshifts seen in the panels of Fig.~\ref{fig:smt_of_HSClike_samples} to not have a very explicit dependence on the true redshift.

We compute the abundance of the \pzmock samples by using the observed stellar masses and redshifts by simply counting the number of galaxies in given observed threshold stellar-mass bins, and dividing by the volume as interpreted based on the observed redshifts. We show the abundances of the Pz- and UM-mock samples as blue and orange lines in Fig.~\ref{fig:abundance} and discuss them in Section~\ref{res:abundance}. 

We have shown that the samples selected in the observed stellar masses and photometric redshifts are thus expected to be related to each other in a non-trivial manner due to the coupling of the stellar mass and redshift errors. This can potentially impact the observables of our interest, namely the measured abundance as well as the WLS. These differences can result in a systematically biased inference of the galaxy-dark matter connection from such measurements.

\section{Weak lensing signal measurement} \label{measurement}
In this section, we describe our methodology to compute the WLS as it would be measured by a mock survey that mimics the HSC given the errors in the photometric redshift and the associated errors in the stellar mass inference. 

For a lens galaxy, the WLS a.k.a Excess Surface Density (ESD), denoted by $\Delta \Sigma (R)$, at a projected radius $R$ from the lens, is a product of the average tangential shear $\langle \gamma_t \rangle (R)$ imparted in the shapes of background galaxies and a geometrical factor, the critical surface density $\Sigma_\rmcrit (\zl,\zs)$, that depends on the lens and source redshift pairs ($\zl$, $\zs$) via angular diameter distances amongst the lens, the source and the observer,   
\begin{align}
    \Delta \Sigma(R) = \langle\gamma_{\rm t}\rangle(R)\,\Sigma_{\rm crit}(\zl,\zs)\,. \label{eq:esd}
\end{align}
As discussed in Section 8 of \citetalias{2024MNRAS.527.5265C}, the WLS measured from a survey with photometric redshifts for the lenses, $\Delta \Sigma_\rmo$, is affected due to both the factors on the right-hand side. Uncertainties in the lens redshifts lead to selection of \lspair{} pairs which are not really separated by the expected projected distance $R$ as demonstrated in Fig.~\ref{fig:demo_figure} and cause mixing of \lspair{} pairs across the radial bins. Moreover, the amplitude of the measured lensing signal gets corrupted due to the associated uncertainty in the geometrical factor. In this work, we account for both of these effects by applying a similar method as developed in \citet{2013ApJ...777L..26M} and implemented in \citet{2015ApJ...806....2M}.

Consider a single lens galaxy with a true weak lensing observable, $\Delta \Sigma_\rmt (R_\rmt)$ at a true projected separation $R_{\rmt}$. Let us consider that its true redshift is $\zt$ and its observed lens photometric redshift $\zo$. The observed WLS due to this galaxy, $\Delta \Sigma_\rmo (R_\rmo)$, will be given by\footnote{$\beta$ can be more or less than 1 even when $\zo>\zt$ depending on the exact values of redshifts and the $P(\zs)$ distribution.},
\begin{align}
    \Delta \Sigma_\rmo (R_\rmo) =\Delta \Sigma_\rmt \left(R_\rmt [R_\rmo]\right) \beta (z_\rmt, z_\rmo, P(z_{\rm s}))\, \label{eq:obsESD}
\end{align}
where,
\begin{align}
    \beta (z_\rmt, z_\rmo, P(z_{\rm s})) = \left[ \frac{\langle \Sigma_{\rmcrit}^{-1}(\zo)\rangle}{\langle \Sigma_{\rmcrit}^{-1}(\zt)\rangle} \right]^{-1}
\end{align}
Here, the angular separation of the source galaxy with the lens galaxy, $\theta$, together with the true (observed) lens redshift can be used to compute the distance $R_\rmt$ ($R_\rmo$), which is the projected comoving separation between the lens and the source galaxy image. Therefore, the relation between $R_{\rmt}$ and $R_{\rm o}$ can be characterised by
\begin{align}
    R_\rmt (R_\rmo) = R_\rmo \times \frac{D_{\rm c}(z_\rmo)}{D_{\rm c}(z_\rmt)}\,,
\end{align}
where $D_{\rm c}(z)$ represents the transverse comoving distance computed at redshift $z$. The multiplicative factor in Eq.~\ref{eq:obsESD}, is given by the ratio of the average of the inverse critical surface densities assuming the true and the observed redshift of the lens. The average of the inverse critical surface densities is carried out over the redshift distributions of the source galaxies, $P(z_{\rm s})$ and is given by
\begin{align}
    \langle \Sigma^{-1}_{\rmcrit}(z) \rangle = \frac{4\pi G (1+z)^2}{c^2} \int_{z}^\infty \frac{D_{\rm a}(z) D_{\rm a}(z, z_{\rm s})}{D_{\rm a}(z_{\rm s})} p(z_{\rm s}) dz_{\rm s} \,. \label{avsigcrit}
\end{align}
With the help of the transformations in Eq.~\ref{eq:obsESD}, we have ensured that the true observable tangential shear $\gamma_{\rm t}(\theta)$  remains unchanged, and our interpretation of this signal as $\Delta\Sigma$ is affected due to our imperfect knowledge of the lens redshift.

The effect of the photometric redshifts for the lens galaxies on the stacked lensing signal for the \pzmock sample can then be computed as the average over galaxies in the lens samples, 
\begin{align}
    \Delta \Sigma_\rmo (R_\rmo) = \frac{\sum_{i} w_{\rm lc}^i(z_\rmt)\, \Delta \Sigma_\rmo^i (R_\rmo)}{\sum_{i} w_{\rm lc}^i(z_\rmt)}\,, \label{eq:stackedWLS_obs}
\end{align}
where $w_{\rm lc}$ is the light cone weight.

We next describe our procedure to compute the $\Delta\Sigma_{\rmt}$ required in Eq.~\ref{eq:obsESD}. In principle, we could compute the true WLS around each UM galaxy by utilising the dark matter distribution around it. However, such an exercise would be quite time-consuming. Therefore, we carry out this exercise statistically, instead. We first obtain the central and satellite HODs from the simulation box (at the median redshift of HSC galaxies in the redshift bin of interest) for true stellar mass ($M_{*, \rmt}$) threshold samples ranging from $\logMstarlimit$ of $7.0$ to $10.6$ in bins of 0.5 dex and then $10.6$ to $11.2$ in bins of $0.2$ dex. The halo occupation distributions are obtained in 40 log-spaced halo mass bins. We have ignored some of the halo mass bins for the HOD computation if the HODs in 4 of the neighbouring mass bins are sparsely populated due to Poisson fluctuations. \footnote{The UM-DR1 catalogue contains the virial mass estimate, $M_{\rm vir}$, while we assume it to be $M_{\rm 200m}$ to remain consistent with our HOD modelling scheme. The difference between the lensing signals due to the use of either halo mass is less than 2 per cent in either redshift bins, smaller than the statistical errors.}. We subtract the HODs of the consecutive threshold samples to obtain the HODs in bins of stellar mass. We show these HODs in Appendix~\ref{app:UT_HODplots}.

We interpolate the HOD in each stellar mass bin on a cubic spline and analytically compute the expectation of the true WLS of such galaxies using the formalism in \citet[][see also Sec.~\ref{theoretical_modelling} below]{2013MNRAS.430..725V}. While computing the stacked expectation of the WLS in Eq.~\ref{eq:stackedWLS_obs}, we use this value of $\Delta\Sigma_{\rmt}$ for every galaxy that falls in the observed redshift and stellar mass bin in Eq.~\ref{eq:obsESD}. Given that we use the HOD of galaxies at the median redshift of the sample, we have effectively ignored any redshift evolution of the HOD within the single bin. However, we do account for the fact that the halo mass, halo bias and the power spectrum of matter vary with redshift, and use the corresponding quantities while computing the expectation of the WLS. For computational efficiency, this is computed using a lookup table of the WLSs as a function of the true redshifts of these galaxies. We also add a contribution due to the baryonic mass component of galaxies in the form of its stellar mass to the true WLS.

In addition to the observed WLS, we also compute the WLS of galaxies if they were binned with their true stellar masses and redshifts (UM-mocks). 
In Fig.~\ref{fig:underlying_vs_obs_wls}, we compare the WLS from a \pzmock sample with that in the UM-sample. The difference between the two WLSs is a result of both, sample selection as well as inaccuracies in assigning projected radii and the critical surface densities due to lens redshift errors. These differences in the WLS can potentially result in differences in the inferred halo occupation distribution parameters which we aim to quantify.

\begin{table}
    \centering 
    \begin{tabular}{llcclc}
        \hline\hline \\[-1.5ex]
        \multicolumn{1}{c}{} & \multicolumn{2}{c}{$ z_\rmo\, \in [0.30, 0.55) \coloneqq z_1$} & & \multicolumn{2}{c}{$z_\rmo\, \in [0.55, 0.80)\coloneqq z_2$} \\[1ex]
        \cline{2-3} \cline{5-6}\\[-2ex]
        \multicolumn{1}{c}{ $\log \left[ \frac{M_{*,\rm lim}}{\hiimsun} \right]$ } & $z_{\rm med}$ & $\log \tilde{M}_{\rm bary}$ & & $z_{\rm med}$ & $\log \tilde{M}_{\rm bary}$ \\[1ex]
        \hline\\[-2.5ex]
        8.6 & 0.45 & 10.06 & &  --- & ---\\
        8.8 & 0.45 & 10.14 & &  --- & ---\\
        9.0 & 0.45 & 10.22 & &  0.69 & 10.20\\
        9.2 & 0.45 & 10.29 & &  0.69 & 10.28\\
        9.4 & 0.45 & 10.37 & &  0.69 & 10.36\\
        9.6 & 0.45 & 10.45 & &  0.69 & 10.43\\
        9.8 & 0.45 & 10.54 & &  0.69 & 10.52\\
        10.0 & 0.45 & 10.65 & &  0.69 & 10.61\\
        10.2 & 0.46 & 10.76 & &  0.69 & 10.72\\
        10.4 & 0.45 & 10.87 & &  0.69 & 10.84\\
        10.6 & 0.45 & 11.05 & &  0.69 & 10.99\\
        10.8 & 0.45 & 11.21 & &  0.69 & 11.15\\
        11.0 & 0.44 & 11.38 & &  0.68 & 11.32\\
        11.2 & --- & --- & &  0.68 & 11.49\\        
        \hline
    \end{tabular}
    \caption{\pzmock sample statistics: The median redshift $z_{\rm med}$ and the average stellar mass $\tilde{M}_{\rm bary}$ of galaxies in units of $\himsun$ in each threshold sample within redshift bins $z_1$ and $z_2$ that we use in our study.}
    \label{tab:z1z2data}
\end{table}

\section{Theoretical modelling} \label{theoretical_modelling}
\subsection{Analytical HOD framework}
In \citetalias{2024MNRAS.527.5265C}, we have used the halo model-based theoretical framework of \citet{2013MNRAS.430..725V} as implemented in, \citet{2015ApJ...806....2M}, to predict the abundance and WLS of the galaxy sample. Given that we have already summarised this model in \citetalias{2024MNRAS.527.5265C}, we only list the ingredients and the key assumptions to predict the observables from this formalism here. We assume that a simple halo-mass based halo occupation distribution (HOD) model with five free parameters \citep{2005ApJ...630....1Z, 2005ApJ...633..791Z}, two for central galaxies ${\Mmin, \sigmalogM}$ and three for satellite galaxies ${M_0, M_1, \alpha}$ can be used to adequately model the observables. The HOD describes the average number of galaxies from the threshold sample living in a halo of mass $M$, $\avg{N}(>M_{*, \rm limit}|M) \equiv \avg{N}(M)$, and is expressed as a sum of two parts. i.e. $\avg{N}(M)=\avg{\nsat}(M)+  \avg{\ncen}(M)$. The average occupation of the central galaxies $\avg{\ncen}(M)$ and the satellite galaxies $\avg{\nsat}(M)$ are expressed as a complementary error function and a modulated power law, respectively. These halo occupation distributions depend upon the model parameters, such that,  
\begin{align}
    \avg{\ncen}(M) =& \frac{1}{2} \left[1+{\rm erf}\left(\frac{\log {M - \log  \Mmin}} {\sigmalogM} \right)\right] \label{cenHOD}, \\
    \avg{\nsat}(M) =& \avg{\ncen}(M) \left( \frac{M-M_0}{M_1} \right)^\alpha. \label{satHOD}
\end{align}

We treat the two halo occupations to be independent of each other and assume that the central galaxies live at the centre of their dark halos, while the satellite galaxies are distributed according to the same NFW profile as the matter distribution within the host halo. Given that subhalo mass fractions are not larger than 10 per cent \citep{2011MNRAS.410.2309G, 2016MNRAS.458.2870V} of the total mass of the parent halo, we have ignored the contribution of subhalos which host the satellite galaxies. For modelling purposes, we define halos to enclose a matter overdensity which is 200 times the mean matter density of the Universe and use the corresponding mass $M_{\rm 200m}$ and concentration $c_{\rm 200m}$ while computing the model predictions. We assume the concentration-mass relation of \citet{2007MNRAS.378...55M} as the mean relation and allow for a scaling freedom in the relation by using a multiplicative free parameter $c_{\rm fac}$ to marginalise over the processes affecting the concentration of dark matter halos at a fixed halo mass. 

The HOD framework outputs the WLS due to the matter distribution associated with the lenses for a set of HOD parameters assuming it follows the expectation from collisionless simulations, but in the inner regions, the observed lensing signal can be affected by the baryonic content as well. Therefore, our model also includes the contribution of these baryons as a central point-mass contribution, 
\begin{align}
   \Delta \Sigma_{b}(R) = \frac{\bar{M}_{\rm bary}}{\pi R^2} \label{esd_stelm}\,.
\end{align}
We have shown in \citetalias{2024MNRAS.527.5265C} that the contribution of the baryonic component to the WLS for $R>100\chikp$ increases with the stellar mass threshold but is less than $10$ per cent even for the most massive threshold sample (see dashed black line in Figs.~5 and 6 in paper I). Therefore we restrict our signal measurements within the same regime. The list of average stellar masses of galaxies used as $\tilde{M}_{\rm bary}$ in units of $\himsun$ in the mock and true threshold samples are listed in Table~\ref{tab:z1z2data}.

The abundance of galaxies for each threshold sample is another observable and can be computed using the HOD as
\begin{eqnarray}
n_{\rm gal} = \int \rmd M\,n(M) \avg{N(M)}\,,
\end{eqnarray}
where $n(M)$ is the halo mass function \citep{2010ApJ...724..878T}. In \citetalias{2024MNRAS.527.5265C}, we showed that the Bayesian analysis results in quite a degenerate inference in terms of the standard halo occupation distribution parameters which are used to describe our model. However, we also showed that the Bayesian analysis robustly constraints the average lensing masses for the halos of central lens galaxies $\avMcen$ of the sample which can be computed from the HOD as,
\begin{eqnarray}
\avg{M_{\rm cen}} = \frac{\int \rmd M\,M\,n(M) \avg{\ncen(M)}}{\int \rmd M\,n(M) \avg{\ncen(M) }}  \label{eq:avMcen}\,.
\end{eqnarray}
The fraction of satellite galaxies in a given stellar mass threshold bin is another inferred quantity from the HOD and is given by,
\begin{eqnarray}
\fsat = \frac{\int \rmd M\,n(M) \avg{\nsat(M)}}{ \int \rmd M\,n(M) \avg{N(M)} }\,. \label{eq:satfrac}
\end{eqnarray}
The HOD modelling scheme assumes the median redshift of the sample to compute the observables, and the assumed functional form corresponds to true stellar mass threshold samples, rather than the complex selection that is a result of the errors in the redshifts of our lenses. The comparison of our modelling results from the \pzmock with the underlying truth in the UM-mock will allow us to quantify the biases that may arise.

\subsection{Covariance matrix}
Obtaining the posterior distribution of parameters given a measured observed signal requires comparing the signal to the model prediction and the covariance matrix provides the metric to be used in this case. The errors on the measured WLS in \citetalias{2024MNRAS.527.5265C} were dominated by the intrinsic distribution of the shapes of galaxies, also called shape noise. The shape noise depends on the number of \lspair{} pairs used to compute the WLS at a given projected radial bin and the use of the same source galaxy for the measurement of the signal around two different lens galaxies at different projected separations, can correlate the errors in different radial bins. On large scales, the coherent large-scale structure also induces covariance which we had captured by the jackknife method in \citetalias{2024MNRAS.527.5265C}. 

Our methodology to compute the mean lensing signal from observations results in systematic shifts in the underlying signal due to the errors in the photometric redshifts of the lens galaxies, when compared to using the true lens redshifts for the sample selection and signal measurement. Given that we would like to gauge the impact of these systematic shifts on the inferred halo mass measurements, we will model noiseless data vectors, but with the same covariance matrix as we used in our analysis in \citetalias{2024MNRAS.527.5265C}. Any systematic shift in the inferred halo-mass stellar-mass scaling relation compared to the underlying truth can then be estimated given the expected statistical errors. We expect these systematic biases to be dominated by sample selection issues and the inaccuracies in the redshift, although model misspecification biases could also play a role. We examine the model misspecification bias separately in Appendix~\ref{app:modelbias}.

\subsection{Bayesian analysis}
We probe the posterior distribution of the model parameters using the same Bayes theorem implementation framework discussed in Section 5, \citetalias{2024MNRAS.527.5265C}. The modelling prescription in \citetalias{2024MNRAS.527.5265C} uses a broad range of uninformative priors and realistically measured errors and covariance matrices. In this work also, we use the same set of priors for the free parameters as listed in Table~\ref{tab:priors}, the same covariance matrices for the lensing signals and a relative error of 15 per cent on the galaxy abundance estimates for each stellar mass threshold sample in order to fit the measurements for both UM and \pzmock samples. 
\begin{table}
    \centering 
    \renewcommand{\arraystretch}{1.2}
    \begin{tabular}{cl}
    \hline
         Parameters &  Priors \\
         \hline
         $\log \left[ M_{\rm min}/ (\himsun) \right]$ & flat $(10.0, 16.0]$  \\
         $\log \left[ M_1/ (\himsun) \right]$ & flat $(10.0, 16.0]$ \\ 
         $\log \left[ M_0/ (\himsun) \right]$ & flat $(6.0, 16.0]$ \\
         $\alpha$ & flat (0.001, 5.0] \\
         $\sigmalogM$ & flat (0.001, 5.0] \\
         $c_{\rm fac}$ & $G(\mu=1,\sigma=0.2), >0$ \\ 
         \hline
    \end{tabular}
    \caption{Priors for our model parameters: The same priors have been used for all the stellar mass threshold samples in both redshift bins. All the flat prior ranges are wide enough to keep them uninformative.}
    \label{tab:priors}
\end{table}

The abundance and galaxy-galaxy lensing signals given the HOD parameters and model ingredients are predicted with the help of the analytical HOD modelling software {\sc aum} \citep{2021ascl.soft08002M}. We have used the the publicly available package {\sc emcee v3.1.1} \citep{2013PASP..125..306F} which is an affine invariant MCMC ensemble sampler based on \citet{2010CAMCS...5...65G} to sample from the posterior distributions of our model parameters. We run the chains with 128 walkers for a total of 10000 steps and remove the first 2000 steps from each walker as a burn-in phase after verifying the stationarity of our parameters of interest to confirm convergence. 

\section{Results} \label{results}

As results, we now present the comparison of our analysis of the \pzmock with the underlying truth in the UM-mock. In Section~\ref{subsec:impact}, we will first demonstrate the impact of the photometric redshift errors on the observables of our interest, namely the abundance and the WLS. In Section~\ref{subsec:hod_impact}, we will present the results of the HOD modelling of these observables to assess the systematic impact on the inferred scaling relation between stellar mass and halo mass, and the satellite fractions inferred from our analysis.

\subsection{Impact of lens redshift errors on observables}
\label{subsec:impact}
\subsubsection{Abundance of galaxies} 
\label{res:abundance}
\begin{figure}
    \centering
    \includegraphics[width=\columnwidth]{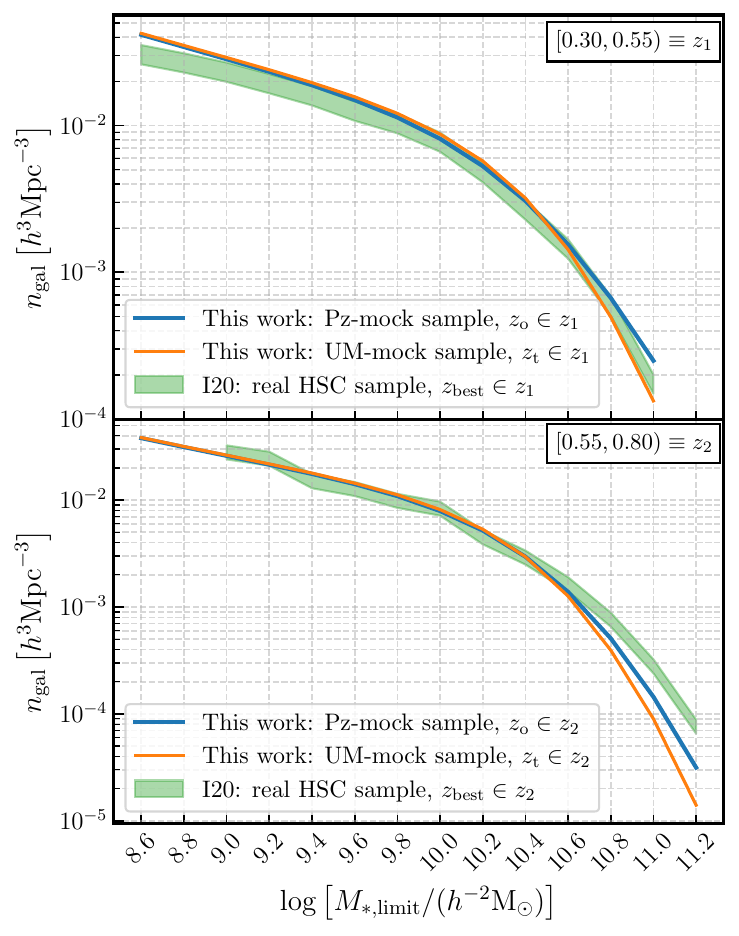}
    \caption{Abundances of \pzmock (blue) and UM-mock (orange) stellar mass threshold samples constructed in this work. The circular data points show the abundance measurements of the real HSC photometric samples used in \citetalias{2024MNRAS.527.5265C} for joint modelling along with the WLS. The top and bottom panel correspond to datasets in $z_1$ and $z_2$ bin.}
    \label{fig:abundance}
\end{figure}

The errors in redshift of the lens galaxies result in changes to the sample of galaxies selected in various observed stellar mass thresholds and redshift bins. Therefore, we expect that the abundance of galaxies can be systematically affected by these errors. We compare the abundance of the observed sample from the \pzmock (blue solid line) to the underlying true abundance in the UM sample (orange solid line) in Fig.~\ref{fig:abundance}. The two panels correspond to the two redshift bins of our interest, respectively. We observe that in both bins, the abundance of galaxies in the stellar mass threshold samples with $\logMstarlimit <10.6$ differ very little from the true underlying abundance (a difference of less than 10 per cent). 

The small difference in the abundance of the \pzmock galaxies compared to UM-mock is related to an approximate cancellation of the number of galaxies that scatter in and out of the sample (see Fig.~\ref{fig:smt_of_HSClike_samples}) in such threshold bins. The differences, however, start to grow larger beyond the characteristic knee of the stellar mass function and the abundance can be overestimated at the high stellar mass threshold bins (as high as 90 per cent for the highest threshold bin in redshift bin 1). This is related to the exponential drop in the number density of galaxies beyond the knee, where there are exponentially fewer higher stellar mass galaxies that scatter out of the sample, compared to the larger sample of lower mass galaxies that scatter in, thus increasing the abundance. This qualitative trend is also the same for the second redshift bin, where the differences in the abundance are fairly small at the low mass end but grow to about 60 per cent at the highest threshold bin.

We also compare the observed abundances to those inferred from the actual HSC survey by I20 using the $\zbest$ estimate of the redshift from HSC. In each panel of Fig.~\ref{fig:abundance}, these measured abundances are shown as green-shaded regions which correspond to the abundances and the 15 per cent errors that we assign in our analyses in \citetalias{2024MNRAS.527.5265C}. This should be compared to the blue solid lines corresponding to the abundance in the \pzmocks{}. Overall given the errors, the abundance of galaxies in the \pzmock has a fairly similar qualitative behaviour as the abundances of galaxies observed in the HSC survey, although the agreement is not entirely quantitatively consistent, and we discuss some of the reasons for this in Section~\ref{challenges}.

\subsubsection{Weak lensing signal of galaxies} \label{impact_lensing_measurement}

\begin{figure*}
    \centering
    \includegraphics[width=\textwidth]{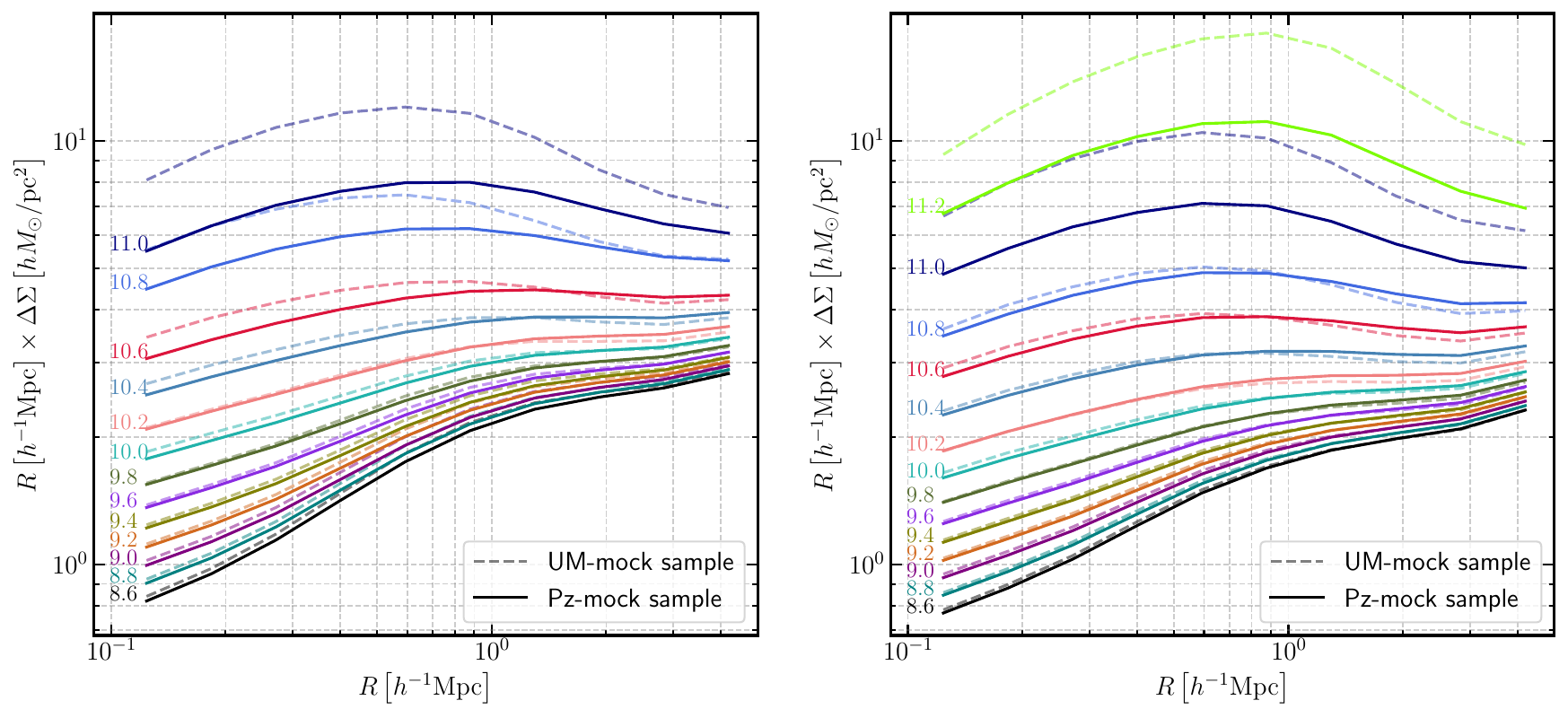}
    \caption{Impact of \photoz{} error on the WLS: The left and right panels show the lensing signal of \pzmock (solid curve) and UM-mock (dashed curve) threshold samples in redshift bins 1 and 2 respectively. Refer to Section~\ref{impact_lensing_measurement} for discussion.}
    \label{fig:underlying_vs_obs_wls}
\end{figure*}

The differences in the sample selection due to redshift errors of lenses, as well as the conversion of the tangential shear to the excess surface density signal and its assignment to a projected radius, affect the measured WLS. The measured WLS of the \pzmock sample for redshift bins 1 and 2 are shown as solid lines in the left and the right-hand panel of Fig.~\ref{fig:underlying_vs_obs_wls}, respectively. The various coloured lines depict the different stellar mass threshold samples. The dashed lines of the same colour correspond to the weak lensing in the UM-mock samples.

For redshift bin 1, the measured signal for the lower stellar mass threshold bins is suppressed compared to the UM-mock signal, albeit by less than 4 per cent. We observe subtle changes to the shape of the measured signal compared to the underlying truth where the fractional decrease in the signal is comparatively larger at intermediate distances of around $1 \chimp$ around the galaxies. As we approach threshold samples with stellar masses higher than the knee of the stellar mass function, we start to see much larger decrements (reaching as large as 34 per cent) in the amplitude of the WLS in the first redshift bin, with further changes to the shape of the signal, and the highest decrement occurring at smaller distances. The changes to the shape of the WLS likely reflect the increased satellite fractions in the observed samples. For redshift bin 2 (right-hand panel of Fig.~\ref{fig:underlying_vs_obs_wls}), the effect is even milder at the lower threshold bins and shows a similar level of decrease in the signal in the higher threshold mass bins as seen in redshift bin 1.

The decrease in the amplitude of the overall signal is related to the contamination from lower stellar mass galaxies in the sample. This overall contamination increases for samples with increasing stellar mass threshold beyond the knee of the SMF and impacts the amplitude of the signal. The effects due to radial mixing and critical surface density changes are comparatively smaller for these samples but not entirely insignificant and both sample selection and lensing measurement effects are important to explain the differences in the observables we see.

\begin{figure*}
    \centering
    \includegraphics[width=\textwidth]{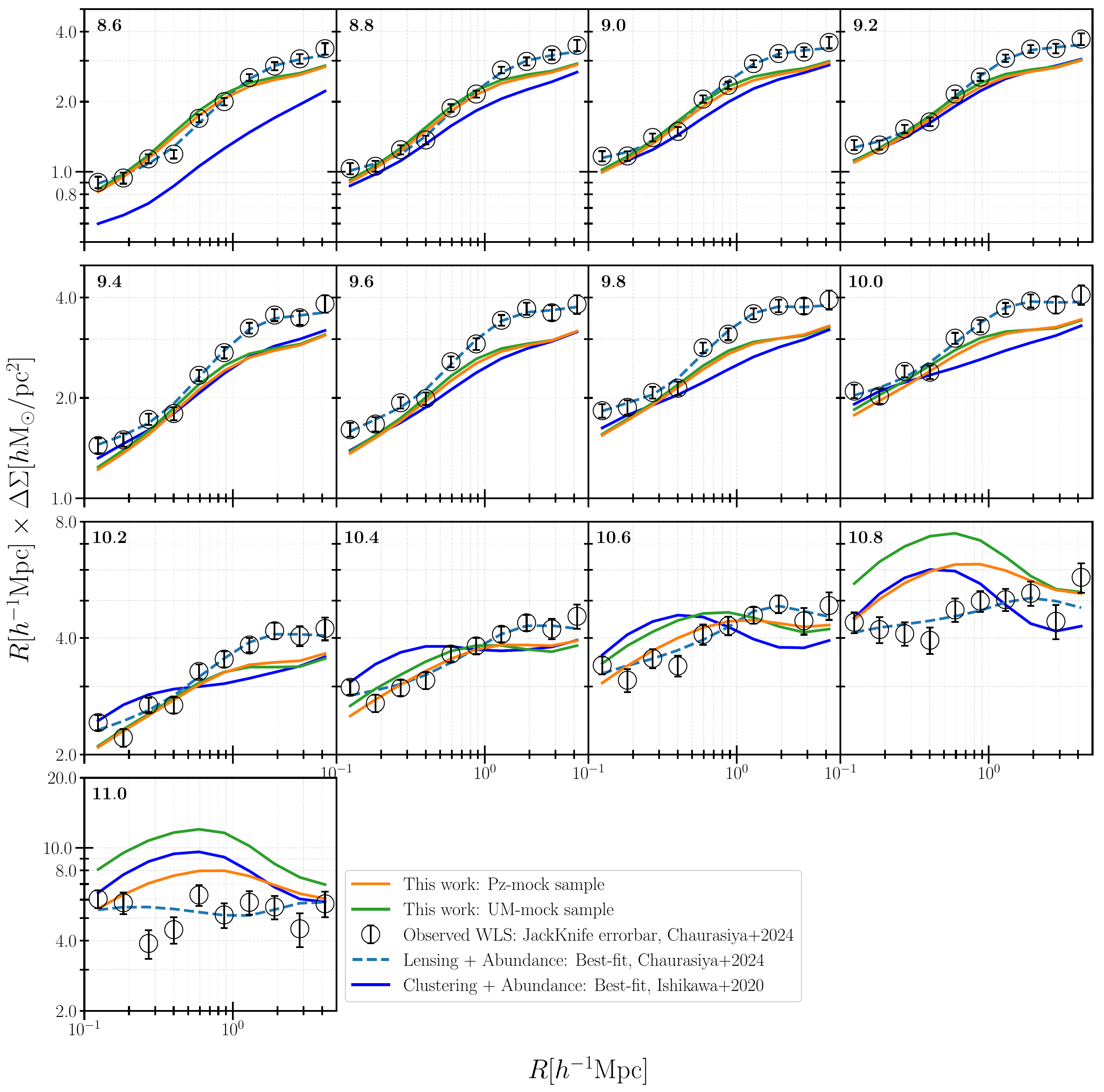}
    \caption{$z_1$ bin: Comparison between the WLS obtained for the various \pzmock stellar mass threshold ($\logMstarlimit$) samples (orange curves) and the WLS measured from the real HSC galaxies in \citetalias{2024MNRAS.527.5265C} (point with error bar). Each panel corresponds to a different $\logMstarlimit$ sample denoted at the top left corner. The data points with error bars in each panel show the signal measurements from \citetalias{2024MNRAS.527.5265C} while the blue dashed curve shows the best-fitting model based on joint fitting of abundance and lensing from there. We also show the signals obtained for the UM-mock samples (green curves) to demonstrate the impact of \photoz{} errors in transforming the green curves into the orange curves. The solid blue curves show the signal predictions from I20 based on HOD modelling of abundance and clustering of the same HSC galaxies used in \citetalias{2024MNRAS.527.5265C}. Refer to Section~\ref{impact_lensing_measurement} for discussion.}
    \label{fig:z1_comparison_wls_HSCreal_obs_UT}
\end{figure*}

We compare the observed WLSs in the \pzmock (orange line) with the real measurements in the HSC survey (points with error bars) in Fig.~\ref{fig:z1_comparison_wls_HSCreal_obs_UT}. For further comparison, we also show the weak lensing signal predicted from the best-fitting HOD model of abundance and the angular clustering signal of HSC galaxies from \citet{2020ApJ...904..128I} as the blue solid line, while the prediction of the HOD model that best fits the weak lensing signal and the abundance measurements of the same galaxies from \citetalias{2024MNRAS.527.5265C} as the blue dashed line. Overall, we observe that the observed WLSs from the \pzmock are a better match to the observed WLSs in HSC, than the clustering and abundance based HOD model prediction. Although we also see qualitative agreement in the shape of the signal in most cases, we do observe some quantitative differences.

Therefore, rather than comparing the stellar mass-halo mass relation obtained directly from the \pzmock with that inferred from the analysis of the real HSC signal, we will model the \pzmock sample and compare the inferred relation to the underlying truth from UM-mock. This will allow us to infer the systematic biases that we might expect on the inferences of galaxy-dark matter connection from the real HSC analysis.

\subsection{Impact on the inferred Galaxy-Dark Matter connection}
\label{subsec:hod_impact}

\begin{figure*}
    \centering
    \includegraphics[width=\textwidth]{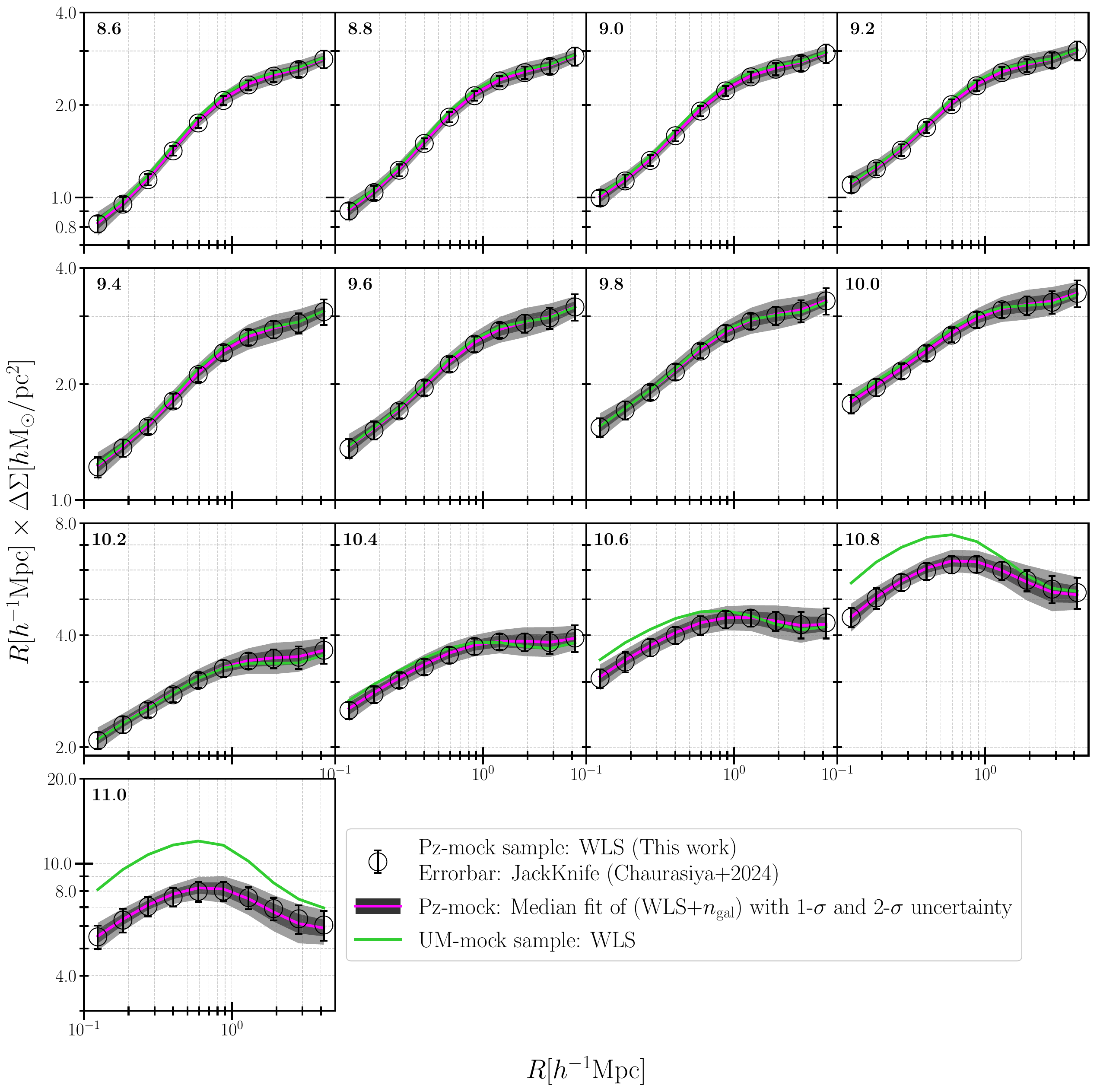}
    \caption{HOD model fitting to the WLS and abundance of \pzmock samples in $z_1$ bin: Each panel corresponds to a particular stellar mass threshold sample indicated at the top left corner. In each panel, the circles with error bars represent the WLS data vector being fit for which the magenta line and the shaded regions around it represent the median fit along with its $1\sigma$ and $2\sigma$ confidence intervals. For comparison, we have also plotted the WLS obtained from the corresponding UM-mock sample as a green line.}
    \label{fig:mcmc_model_z1_HSClike}
\end{figure*}

We carry out a Bayesian inference of the model parameters given the noiseless data vectors\footnote{The data vectors are assumed to be noiseless, but contain systematic shifts compared to the signals in the UM-mock sample, as we saw in Section~\ref{subsec:impact}.} of the observed weak lensing and the abundance of lens galaxies in the \pzmock stellar mass threshold samples, assuming the error covariance matrix from the real HSC survey. The result of this Bayesian analysis which was carried out with the same priors as used for the real HSC analysis is shown for the redshift bin $z_1$ in Fig.~\ref{fig:mcmc_model_z1_HSClike}.

Each panel in Fig.~\ref{fig:mcmc_model_z1_HSClike} shows the $R\Delta\Sigma$ as a function of projected radius $R$ and contains the $1\sigma$ and $2\sigma$ credible intervals shown as shaded regions. We see that, regardless of all the complications regarding sample selection and the WLS measurement present in the threshold samples of the \pzmock, the simplistic HOD model has enough flexibility to fit the observed lensing signal. A good model fit is thus not a guarantee of the absence of any biases. 

In \citetalias{2024MNRAS.527.5265C} we have shown that the posterior distribution of the central and satellite HOD parameters are quite susceptible to the details of the exact abundance of galaxies used to infer these parameters, and that these parameters are highly degenerate with each other. We also showed that the inferred credible intervals on the derived parameter $\avMcen$, which is the average of the halo masses of the central galaxies in the stellar mass threshold bin, are much less susceptible to the differences in these inputs. This derived parameter is a robust combination of the strong degeneracy between the central HOD parameters $\logMmin$ and $\sigmalogM$ which is sensitive to weak lensing. Therefore, in the left panel of Fig.~\ref{fig:z1_HSClike_vs_UT_avMcen_fsat}, we show $\avMcen$ as a function of the stellar mass threshold as blue points with errors for \pzmocks{} in the redshift bin 1. This can be compared to the underlying truth in the UM-mocks. 

Consistent with the small differences in the weak lensing as well as abundance signal between the \pzmock and UM-mock for the low stellar mass threshold bins, we see little to no systematic bias for stellar mass thresholds $\logMstarlimit<10.6$. For higher stellar mass threshold bins, we see that the $\avMcen$ may be systematically biased up to 0.2 dex, corresponding to a systematic deviation which is about $2\sigma$ downwards. Given the relatively low impact of the abundance measurements on the inference of $\avMcen$ as shown in \citetalias{2024MNRAS.527.5265C}, these systematics are likely related to underlying differences in the lens galaxy samples between UM-mock and the \pzmock for such high stellar mass bin. Similar conclusions can be drawn for the inferred galaxy-dark matter connection in redshift bin 2, which is shown in Fig.~\ref{fig:z2_HSClike_vs_UT_avMcen_fsat}.

The true HOD of galaxies even in the true UM-mock does not have the assumed functional form of the HOD that was used for the modelling. Thus any differences in the inference of the galaxy-dark matter connection from our exercise above could also be partly a result of model mis-specification bias. Therefore, we have also modelled the true signals from the UM-mock sample of galaxies with the simplistic HOD model with similar success. The results of this exercise are shown with the orange credible intervals in Fig.~\ref{fig:z1_HSClike_vs_UT_avMcen_fsat} and \ref{fig:z2_HSClike_vs_UT_avMcen_fsat} and these should be compared to the green symbols which shows the underlying truth in the UM-mock. Given the excellent agreement between the green points and the orange credible intervals in both figures, we conclude that model misspecification bias is a much smaller effect.

\subsubsection{Satellite fraction}    
The satellite fractions inferred based on the HOD modelling of abundance and WLS of Pz- and UM-mock galaxies in these samples are shown as blue points with errors and shaded regions in the right-hand panel of Figs.~\ref{fig:z1_HSClike_vs_UT_avMcen_fsat} and~\ref{fig:z2_HSClike_vs_UT_avMcen_fsat} for the two redshift bins respectively. We see that these constraints agree with each other owing to the large errors. Also, the credible intervals for UM-mock samples are consistent with the true satellite fractions in the UM-mocks shown as green points. Despite the small differences in satellite fractions as a function of redshift in different mocks (UM and Pz), the inferred satellite fractions from the HOD model show no sign of redshift evolution.  

\subsubsection{Impact on redshift evolution of galaxy-halo connection} 
\label{redshift_evolution}

We have demonstrated in \citetalias{2024MNRAS.527.5265C} that systematic biases in abundance estimates can indeed impact our ability to infer the evolution in galaxy-halo connection relations. In the present study, similar abundance changes are also seen due to \photoz{} errors along with an additional bias in the lensing signal. We present a comparison of the evolution of average central-halo mass as a function of the stellar mass threshold in UM- and \pzmock samples after accounting for the HSC \photoz{} errors in Fig.~\ref{fig:z1t2_HSClike_vs_UT_avMcen}. The $\avMcen$ constraints of UM-mock samples shown as the blue and orange shaded regions for the redshift bins $z_1$ and $z_2$ respectively, separate out at the $1\sigma$ level beyond the threshold of $\geq 10.6$. While our analysis of \pzmock samples, blue and orange points with error bars for $z_1$ and $z_2$ bins respectively, show that the \photoz{} errors play a role so as to wash out such differences across redshift bins. Whereas for the lower thresholds, the halo masses are consistent within $1\sigma$ in both the redshift bins and \photoz{} errors do not seem to have any significant impact on \pzmock relations compared to the simulation truth. These trends stand as an interesting parallel with the observation made in Fig.~12 of \citetalias{2024MNRAS.527.5265C} that even a similar systematic bias in abundance estimates can eliminate the signs of redshift evolution inferred up to $1\sigma$ significance. Thus the redshift evolution of the high mass slope of the stellar mass-halo mass relation is not only sensitive to the weak lensing but also to the systematic biases in abundance estimates. Therefore to probe any evidence of redshift evolution of galaxy-halo connection, even the higher order effects not accounted for in this study (such as HOD evolution in modelling) can play an important role.

\subsubsection{Implications for average central halo masses inferred from HSC photometric galaxies}
\label{real_galhalo_connection_systematics}
\begin{figure*}
    \centering
    \includegraphics[width=\textwidth]{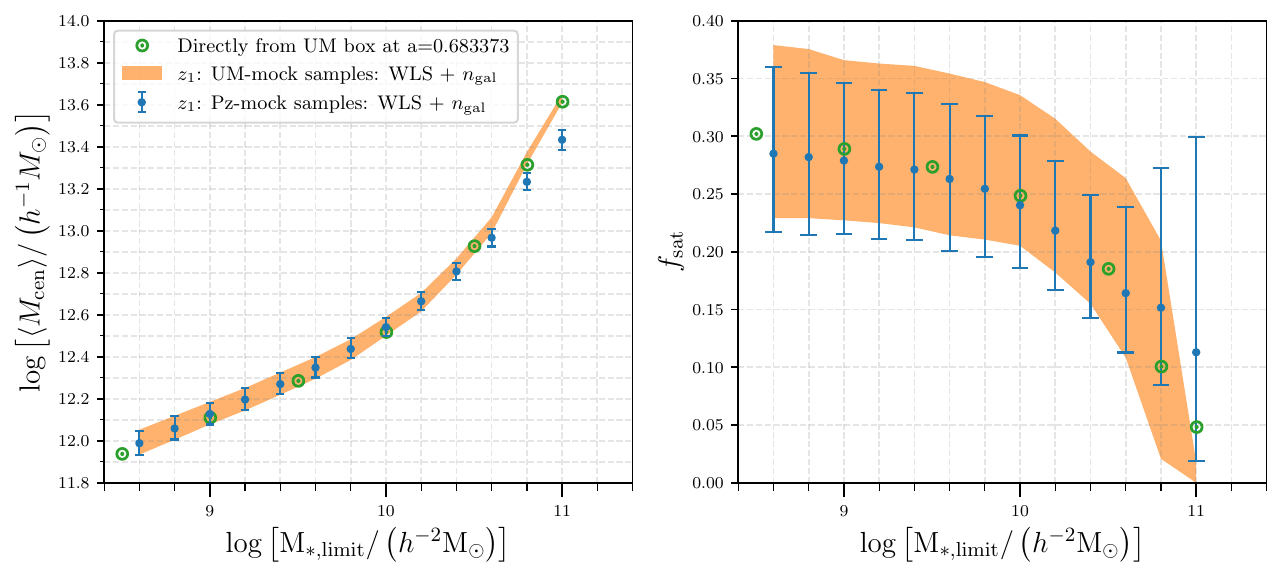}
    \caption{Constrains on average halo mass $\avMcen$ (left panel) and satellite fraction $\fsat$ (right panel) as a function of the stellar mass threshold $\logMstarlimit$ in redshift bin $z_1$: The blue points with errors indicate the $1\sigma$ constraints from fitting the WLS and abundance of \pzmocks{}. This should be compared to the green points which represent the true relation in the the UM galaxy catalogue at the scale factor of 0.683373. The orange band shows similar $1\sigma$ constraints from the UM-mock samples and highlights the success of HOD modelling to capture the underlying true relation in the simulation shown by the green points.}
    \label{fig:z1_HSClike_vs_UT_avMcen_fsat}
\end{figure*}

\begin{figure*}
    \centering
    \includegraphics[width=\textwidth]{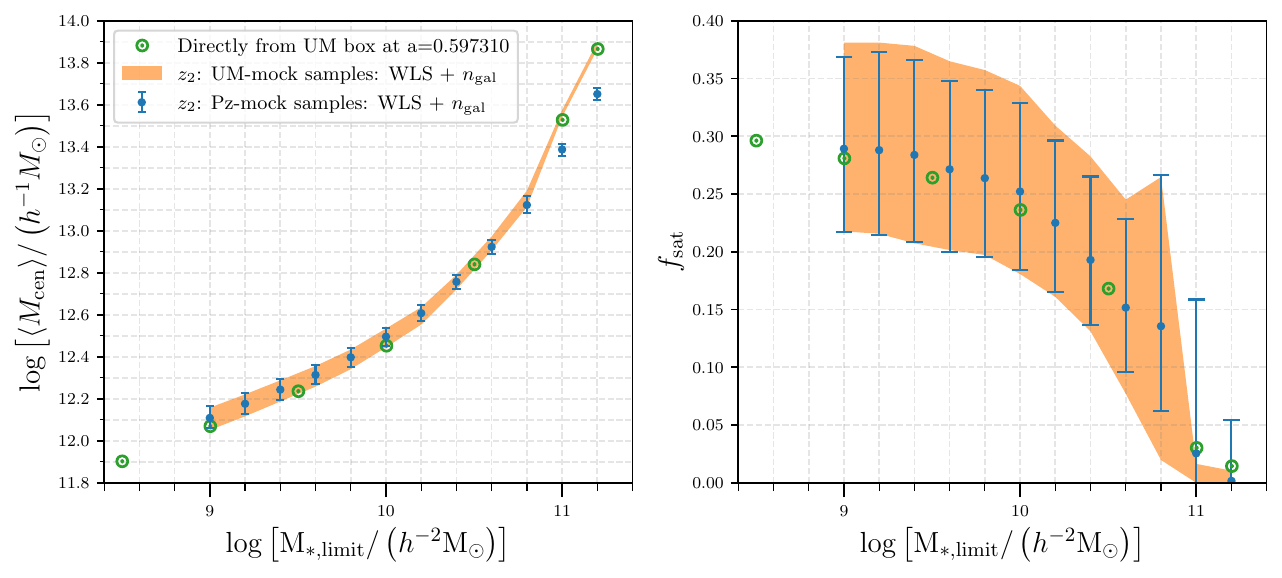}
    \caption{Same as Fig.~\ref{fig:z1_HSClike_vs_UT_avMcen_fsat}, except for redshift bin $z_2$.}
    \label{fig:z2_HSClike_vs_UT_avMcen_fsat}
\end{figure*}

\begin{figure}
    \centering
    \includegraphics[width=\columnwidth]{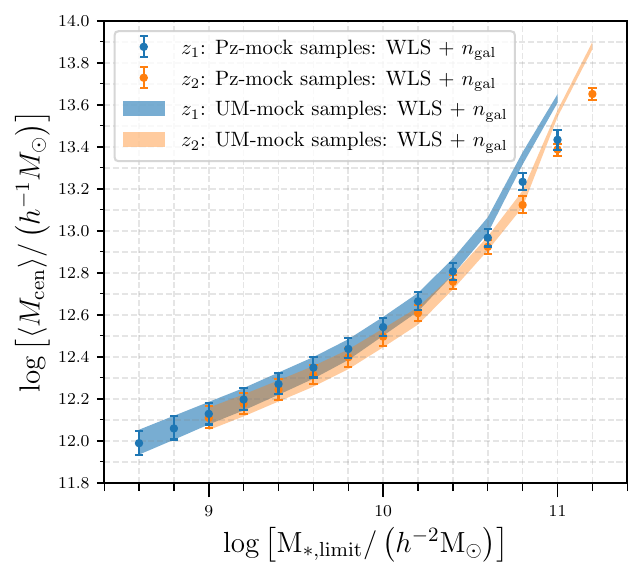}
    \caption{The redshift evolution of average halo masses $\avMcen$ hosting the central galaxies in various stellar mass threshold $\logMstarlimit$ samples across the two redshift bins probed in this work: The bands represent the $1\sigma$ constraints from the UM-mock samples while the points with error bars represent $1\sigma$ constraints from the \pzmock samples. The two redshift bins $z_1$ and $z_2$ are shown in blue and orange colours respectively. Refer to Section~\ref{redshift_evolution} for discussion.}
    \label{fig:z1t2_HSClike_vs_UT_avMcen}
\end{figure}

\begin{figure}
    \centering
    \includegraphics[width=\columnwidth]{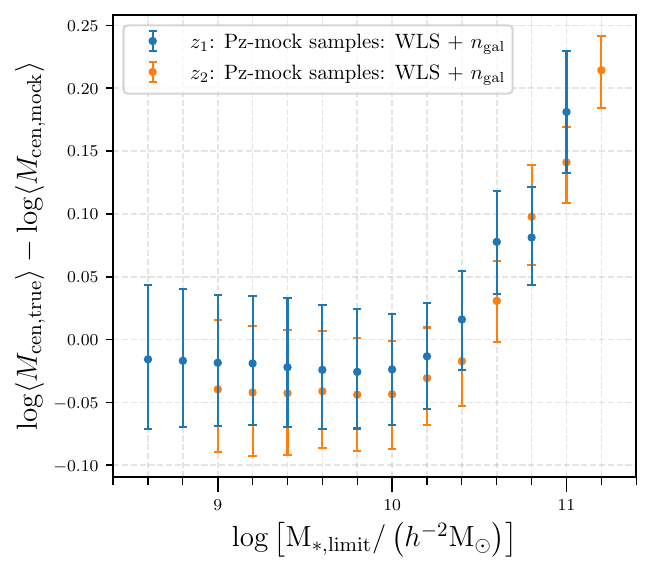}
    \caption{Systematics in the inferred average halo mass $\avMcen$ as a function of stellar mass threshold $\logMstarlimit$ from weak lensing and abundance modelling of the \pzmock photometric galaxy mocks. The blue and orange points correspond to redshift bins 1 and 2 respectively. The systematic shift is computed as the difference between log of the true average halo mass in the UM catalogue $\log \langle M_{\rm cen, true} \rangle $ as calculated from the true HOD (see Fig.~\ref{fig:UTHODs}) using equation~(\ref{eq:avMcen}) and the same obtained from the joint HOD modelling of abundance and lensing of the \pzmock mock samples. The error on each point represents the central 68 per cent confidence interval on the average central halo mass of the mock samples $\log \langle M_{\rm cen, mock} \rangle $. Refer to Section~\ref{real_galhalo_connection_systematics} for discussion.}
    \label{fig:z1t2_HSClike_vs_UT_avMcen_difference}
\end{figure}

The constraints on average central halo mass $\avMcen$ as a function of stellar mass threshold $\logMstarlimit$ derived in \citetalias{2024MNRAS.527.5265C} for the galaxies in the HSC survey are prone to systematics due to complicated interplay of the \photoz{} errors in the population of lens galaxies employed in the analysis. In this work, we have estimated the level of potential systematic shifts in these halo masses after accounting for the expected \photoz{} errors in the HSC survey. We compute the difference between log of the true average halo mass, $\log \avg{M_{\rm cen, true}}$, as computed from the threshold HODs of UM-mocks (also shown as green circles in the left-hand panels of Figs.~\ref{fig:z1_HSClike_vs_UT_avMcen_fsat} and \ref{fig:z2_HSClike_vs_UT_avMcen_fsat}) and log of the median constraint on the corresponding halo mass, $\log \avg{M_{\rm cen, mock}}$, obtained from the joint modelling of abundance and WLS of the \pzmocks{} as shown in Fig.~\ref{fig:z1t2_HSClike_vs_UT_avMcen_difference} and the corresponding central 68 per cent confidence interval is put as the error bar on each point. The blue and orange denote the redshift bins 1 and 2 respectively.

Our analysis indicates that the inferred halo masses in low threshold samples up to 10.4 are consistent with no systematics within the errors in both the redshift bins, however these halo masses in redshift bin 2 show tendency to be slightly overestimated compared to the truth. The analysis of thresholds higher than 10.4, although limited by the size of the redshift calibration sample in the COSMOS field, shows evidence of underestimation ranging from $1\sigma - 2\, \sigma$ depending on the threshold sample. 
These differences act as upper limits of systematics in the high mass thresholds. However, with the availability of a bigger redshift calibration sample which could also be a better representative of the actual photometric galaxies, these systematics can be better probed. It is expected that the actual \photoz{} errors in massive galaxies could be overestimated in this work. Nevertheless, this remains a case of investigation for a future analysis. 

\section{Caveats and possible future improvements} \label{challenges}
In this work, we have investigated the systematics in the inference of the galaxy-dark matter connection based on the use of lens galaxies with photometric redshifts. We have demonstrated the impact of \photoz{} errors on sample selection, lensing, and abundance measurements and how it affects the inference of the galaxy-halo connection.

In order to characterise the performance of the photometric redshifts in HSC, we have used the common galaxies present in the COSMOS field which have more accurate redshift estimates from the 30-band photometric redshift catalogue. However, given the limited sky area of the COSMOS field, our determination of the redshift performance is subject to sample variance. 
We have assessed the performance of the photometric redshifts of the sample as a whole. In reality, however, the performance is expected to be a function of the apparent flux of the galaxy due to statistical errors in the photometry, as well as the type of the galaxy. 

In our analysis, we have further assumed that the errors in redshifts and the stellar masses are perfectly correlated. Although we expect this to be a dominating factor, differences in the SEDs could reduce this correlation. 
Accounting for any remaining extra contribution to the stellar mass errors could improve the assessment of the systematic effects further. For template fitting methods for the inference of the galaxy properties, it may be useful to output the full error covariance matrix between stellar mass and redshifts, which will allow a proper propagation of the errors.

Alternatively, one could also compute the HSC photometry for UM galaxies based on the SEDs of these galaxies, and include realistic photometric errors depending upon the magnitude of galaxies. The resultant magnitudes together with the errors can then be fed to a photometric redshift template code such as {\sc Mizuki} to obtain the inferred stellar mass and the photometric redshift for each galaxy. These galaxies can then be used to define the stellar mass threshold samples in the different redshift bins to assess the impact of the photometric errors on the galaxy-dark matter connection. Although this is time and resource-consuming, such a procedure will fully forward model all the errors in a flux-dependent manner, and thus create a true HSC-mock sample. It is likely that such a mock will also have a better agreement with the abundances observed in the data, especially at the bright end.

In general, the conclusions that we have derived about the possible systematic effects depend upon the particulars of the simulated galaxy catalogue in the UM in two ways. First, the volume of the UM simulation snapshot is about $100$ times smaller than the actual volume probed in the HSC data that we deduce the systematic effects for, could be affected by any sample variance issues. This is especially true in the higher mass end, where the true HOD of the galaxies is quite poorly sampled. Given that photometric redshift errors imply that our high stellar mass thresholds have a lot of contamination from low mass galaxies, this may not be a dominant effect, but one should be mindful of it once the statistical errors become further smaller. Second, the UM galaxy formation model itself has its own uncertainties, which could potentially affect the biases that we infer. However, our framework can be utilised for any other similar model which provides a catalogue of galaxies. 

In addition, given the limited volume of the UM galaxy catalogues, we have also stacked a single snapshot of UM galaxies without recourse to any evolution. The use of a proper light cone could potentially better help understand the relative importance of the HOD evolution in comparison to the \photoz{} errors alone. We intend to explore such effects in the future. As indicated previously, we have implicitly assumed that the effect of any redshift evolution of HOD is substantially smaller than that of the \photoz{} errors while building the mocks.

Finally, we have not discussed the systematics for the standard halo occupation distribution parameters $\Mmin-\sigmalogM$ due to its dependency on the systematics in the abundance and the robustness of the inferred $\avMcen$. We caution the reader against interpreting the magnitude of the systematic we obtain for $\avMcen$ to imply that the other HOD parameters will also be similarly impacted. In reality both these model parameters show larger positive biases than in $\avMcen$. However, the biases in the two parameters are correlated and result in smaller changes in $\avMcen$. We also check in Appendix~\ref{app:modelbias} whether the systematics we derive are a result of any model parameterization bias, i.e., the incompatibility of the functional form of the HOD with the galaxy catalogues in the {\sc UniverseMachine}. 

\section{Summary} \label{summary}
Deep imaging surveys provide statistically large samples of intrinsically faint galaxies to high redshift, and are thus well-suited for studies of galaxy evolution. However, given the broadband photometry that the imaging surveys have to rely on, the redshift uncertainties can be a cause for significant concern about systematic effects. In this work, we have developed a forward modelling framework to assess the impact of redshift errors of photometric galaxy samples which act as lens galaxies in a weak gravitational lensing analysis. Specifically, our method accounts for the two kinds of issues, sample contamination as well as systematics in the measured signal. We have quantified the resulting systematics in the inferred galaxy-halo relations when these \photoz{} errors are not accounted for with a specific example of the HSC survey.

In order to demonstrate the use of our method, we characterised the statistical profile of the redshift errors for our lens galaxy samples obtained from the {\sc Mizuki} photometric catalogue of the HSC survey. We then generated mock galaxy catalogues with such redshift errors starting from the UM-DR1 catalogues taken at the median redshift of the lowest stellar mass threshold sample utilised in \citetalias{2024MNRAS.527.5265C} in each of the two redshift bins $z_1$ and $z_2$. We assumed that the errors in observed stellar masses would be related to the errors in the redshift and assigned observed stellar masses to these galaxies. We generated samples of galaxies with various thresholds of stellar masses within two observed redshift bins. 

We discussed how the intrinsic properties of such \pzmock stellar mass threshold samples deviate from the underlying true samples (UM-mocks) in the simulated universe represented by the best-fitting model of B19. Using these samples we further computed and compared the WLSs of the UM- and the \pzmock samples. We carried out joint modelling of abundance and WLS of each of the UM- and \pzmock samples in a Bayesian framework and the key results and findings of our analyses are summarised below:   

\begin{enumerate}
    \item The samples of lens galaxies in observed redshift bins and in the observed stellar mass thresholds can contain significant contamination from galaxies that lie outside these thresholds. The amount of contamination depends upon the stellar mass threshold and whether it is less or more massive than the characteristic knee of the stellar mass function. Samples on the heavier side of $\logMstarlimit>10.6$ have increasingly larger contamination, while threshold samples below this limit are less affected in a systematic manner.
    \item We have quantified the systematics induced in the abundance and the weak lensing signal of threshold samples. We show that for thresholds below $\logMstarlimit=10.6$, the contamination results in quantitatively small changes to the measurements given the errors.  For higher mass thresholds, we see a significant change in the amplitude of the weak lensing signal, primarily driven by the sample selection.
    \item We find that the average central halo masses, $\avMcen$, derived from a HOD analysis are not expected to show significant evidence of a systematic bias ($<0.05$ dex) for threshold samples below $\logMstarlimit=10.6$. Thus, the galaxy-halo connection result of \citet{2024MNRAS.527.5265C} for such galaxies in the HSC survey is robust to systematics even though the photometric redshifts were employed.
    \item For more massive samples, the inferred $\avMcen$ are likely expected to be biased lower, with a bias increasing with the stellar mass threshold and reaching values as high as $0.2$ dex at the highest threshold, a systematic bias of approximately twice the statistical error.
    \item Although the inferred values of $\avMcen$ seem to appear robust to the photometric redshift uncertainties for the stellar mass threshold bins with $\logMstarlimit>10.6$, the standard HOD parameters $\Mmin$ and $\sigmalogM$ can be potentially biased in a correlated manner, and thus we express caution to interpret the constraints on these parameters and their evolution.
\end{enumerate}
 
When combined with measurements of galaxy abundance and small-scale galaxy clustering, the gravitational lensing signal of galaxies provides important information about the mass-to-light ratios of the dark matter halos in which these galaxies reside. The mass-to-light ratios are important for obtaining cosmological parameters from the joint analyses of these observables. We have shown how the weak lensing measurements can be systematically affected in the presence of redshift and associated stellar mass errors from photometric data, in particular, that such systematic effects may be smaller in the low stellar mass regime. Analyses such as ours need to be carried out to infer the resultant systematic biases on the galaxy observables and the inferred galaxy-halo connection. This can help identify the galaxy samples where the results may not be severely affected by systematics or quantify the systematics, to use as important inputs from small scales in cosmological analyses.

As data from larger imaging surveys such as Euclid and the Rubin LSST looms on the near horizon, we expect to obtain statistically large samples of galaxies appropriate to study the galaxy-dark matter connection spanning many orders of magnitude in stellar mass and a broad range of redshift. Given the statistics, the systematic effects such as the ones described in this paper will become more significant and need to be carefully accounted for with a forward modelling framework such as the one presented in this paper.

\section*{Acknowledgements}
We thank Aseem Paranjape, Masamune Oguri, Anupreeta More for useful discussions on the current project and their comments on the draft version of this paper. We also thank Divya Rana, Amit Kumar, Arman Shafieloo, Ravi Sheth and Priyanka Gawade for useful discussions. NC is thankful for the financial support provided by the University Grants Commission (UGC) of India. We acknowledge the use of Pegasus, the high-performance computing facility at IUCAA. 

We also acknowledge the use of {\sc colossus} package \citep{2018ApJS..239...35D} to convert the $M_{\rm vir}$ masses to $M_{\rm 200m}$. 

\bibliographystyle{mnras}
\bibliography{example} 

\appendix

\section{Observed - stellar mass threshold sample contamination} \label{all_sample_spoilers}
\begin{figure*}
    \centering
    \begin{subfigure}[b]{0.49\textwidth}
        \includegraphics[width=\columnwidth]{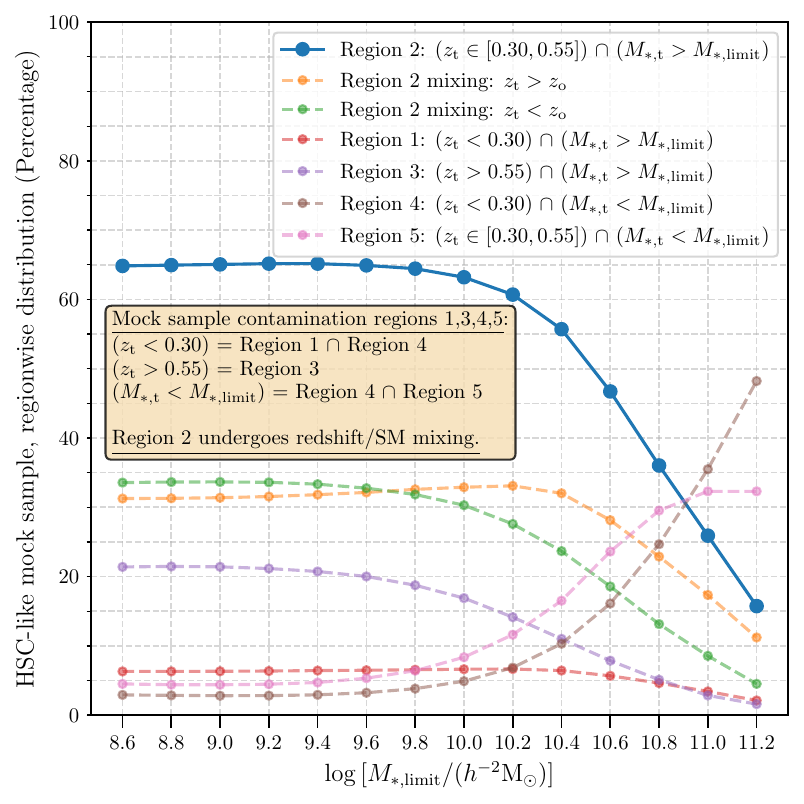}
        \caption{redshift bin $z_{1}$}
        \label{fig:z1_conaminators_and_mixers}
    \end{subfigure}
    \begin{subfigure}[b]{0.49\textwidth}
        \includegraphics[width=\columnwidth]{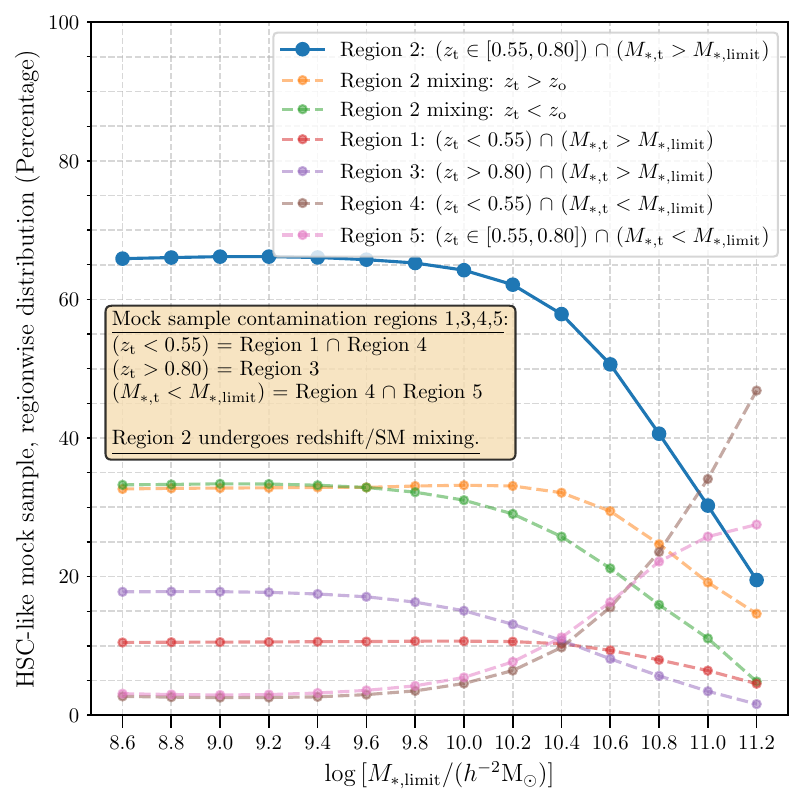}
        \caption{redshift bin $z_{2}$}
        \label{fig:z2_conaminators_and_mixers}
    \end{subfigure}
    \caption{The percentage of mock galaxies occupying region 2 (the sampling space targeted while binning the galaxies) as a function of stellar mass threshold is given by points on the blue solid curve in each panel. The blue curve is divided into two parts based on the inequality between true and observed redshifts of the galaxies in each bin, the points on the green (orange) dashed curve denote the fraction of galaxies that have smaller (larger) true redshifts than the observed ones. While the galaxies excluded by the blue curve denote the sample contamination. This contamination in each threshold bin is further split into regions 1,3,4 and 5 as shown by the various dashed lines in the figures and denoted in the legend. The left and right panels correspond to the redshift bins 1 and 2.}
    \label{fig:conaminators_and_mixers}
\end{figure*}

In Fig.~\ref{fig:smt_of_HSClike_samples}, we had shown how the observed samples are contaminated with galaxies from different regions in the true stellar mass redshift plane. The galaxies occupying region 2 in the left ($z_1$ bin) and right ($z_2$ bin) panels of Fig.~\ref{fig:smt_of_HSClike_samples} are further segregated as galaxies having either a lower or higher observed redshift than their intrinsic truth in Fig.~\ref{fig:conaminators_and_mixers}. The blue curve shows the fraction of galaxies from the full threshold sample which populates region 2 and is further split into the green ($\zo > \zt$) and orange ($\zo < \zt$) dashed curves. The assumed perfect correlation between observed $\Mstar$ and redshift in our mocks implies that the threshold sample 9.8 has a similar number of galaxies in region 2 which either have a lower or a higher $\Mstar$ compared to the intrinsic value, while in the thresholds beyond 10.0, a higher number of (about 10 per cent) galaxies are intrinsically more massive. We note that this trend in region 2 is opposite compared to a large number of intrinsically lower mass contaminant galaxies filling up the threshold samples beyond 10.0.       

The dashed curves corresponding to regions $1$, $3$, $4$ and $5$ collectively represent the total fraction of contaminant galaxies as a function of stellar mass threshold in the mocks. We observe that for the thresholds from $8.6$ to $10.0$, the total contaminant galaxy fraction remains at about $35$ per cent and increases sharply to $80$ per cent for the highest threshold $11.2$ in both the redshift bins. Up to $10.0$, the mocks have slightly more than $50$ per cent galaxies that may have intrinsically higher redshifts and stellar masses which drops to about $15$ per cent for $11.2$ in both the redshift bins.

\section{Intrinsic weak lensing signal of mock-HSC samples} \label{wls_hsclike_vs_intrinsic}
\begin{figure*}
   \centering
   \includegraphics[width=\textwidth]{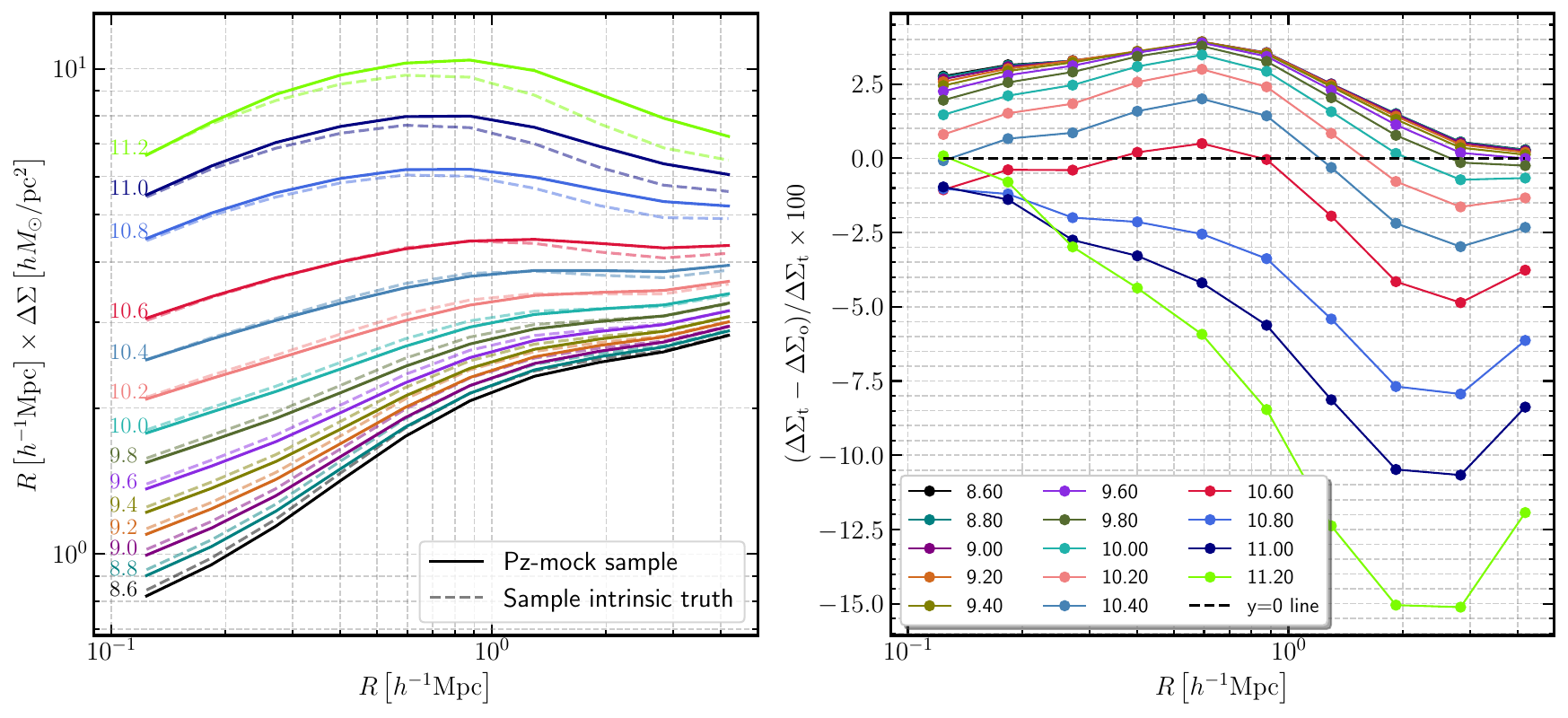}
   \caption{$z_1$ bin: \textbf{(Left panel)} Each pair of a solid and a dashed curve of the same colour represent the WLSs for the same \pzmock stellar mass threshold sample which is indicated on the left side. Each solid curve is computed using the observed redshifts of the \pzmock sample while the corresponding dashed curve represents the intrinsic signal of the same sample computed using the true redshifts. These signals include the effect of both, the dark matter halos and the galaxy stellar masses. \textbf{(Right panel)} The percentage difference between the intrinsic and observed signals for all the threshold samples of the left panel.}
   \label{fig:intrinsic_vs_obs_wls}
\end{figure*}

In Section~\ref{lens_sampling} we have discussed the true properties of galaxies constituting the \pzmock stellar mass threshold samples. Here, we compare the $R \Delta \Sigma$ lensing signal profiles of the \pzmock samples in the redshift bin 1 as a function of the projected separation $R$ in the left panel of Fig.~\ref{fig:intrinsic_vs_obs_wls} in which the profile shown as dashed curve is obtained by using the true redshifts of the galaxies while the other profile shown as the solid curve is obtained using the observed redshifts. The percentage difference between the two estimates of the signal is shown in the right-hand panel which quantifies the impact of the \photoz{} errors via the effects of radial bin mixing and critical surface density scaling.  
Note that a dashed line in Fig.~\ref{fig:intrinsic_vs_obs_wls} can be compared to the corresponding dashed line in Fig.~\ref{fig:underlying_vs_obs_wls} to quantify the effect of sample contamination alone. It turns out that sample contamination plays a dominant role in biasing the lensing signals of massive thresholds.  

We find that in the low mass thresholds below the characteristic knee of the SMF, the halos hosting the galaxies are intrinsically more massive and the observed signals are most suppressed around the distance scales of $0.6\chimp$ by about 4 per cent. For more massive thresholds beyond $10.6$, the observed signal is boosted in general over all the distance scales with maximum effect at the scale of around $2-3\chimp$.     

This comparison also delineates that the galaxies in lower mass threshold samples are not only intrinsically more massive in the stellar component as expected from the Figs.~\ref{fig:diff_z} and \ref{fig:conaminators_and_mixers} but also in the dark component. Similar but opposite inferences can be made for the massive end of the threshold samples.
This finding is non-trivial and for a given depth and completeness of a survey it depends on the interplay among the factors like the slope of the SMF, the survey volume and the \photoz{} error profile as a function of true stellar mass. 

\section{UT threshold samples}
\label{app:utsamples}
\subsection{Halo Occupation Distribution from UM-mock} \label{app:UT_HODplots}
\begin{figure*}
    \begin{subfigure}[b]{0.49\textwidth}
        \includegraphics[width=\columnwidth]{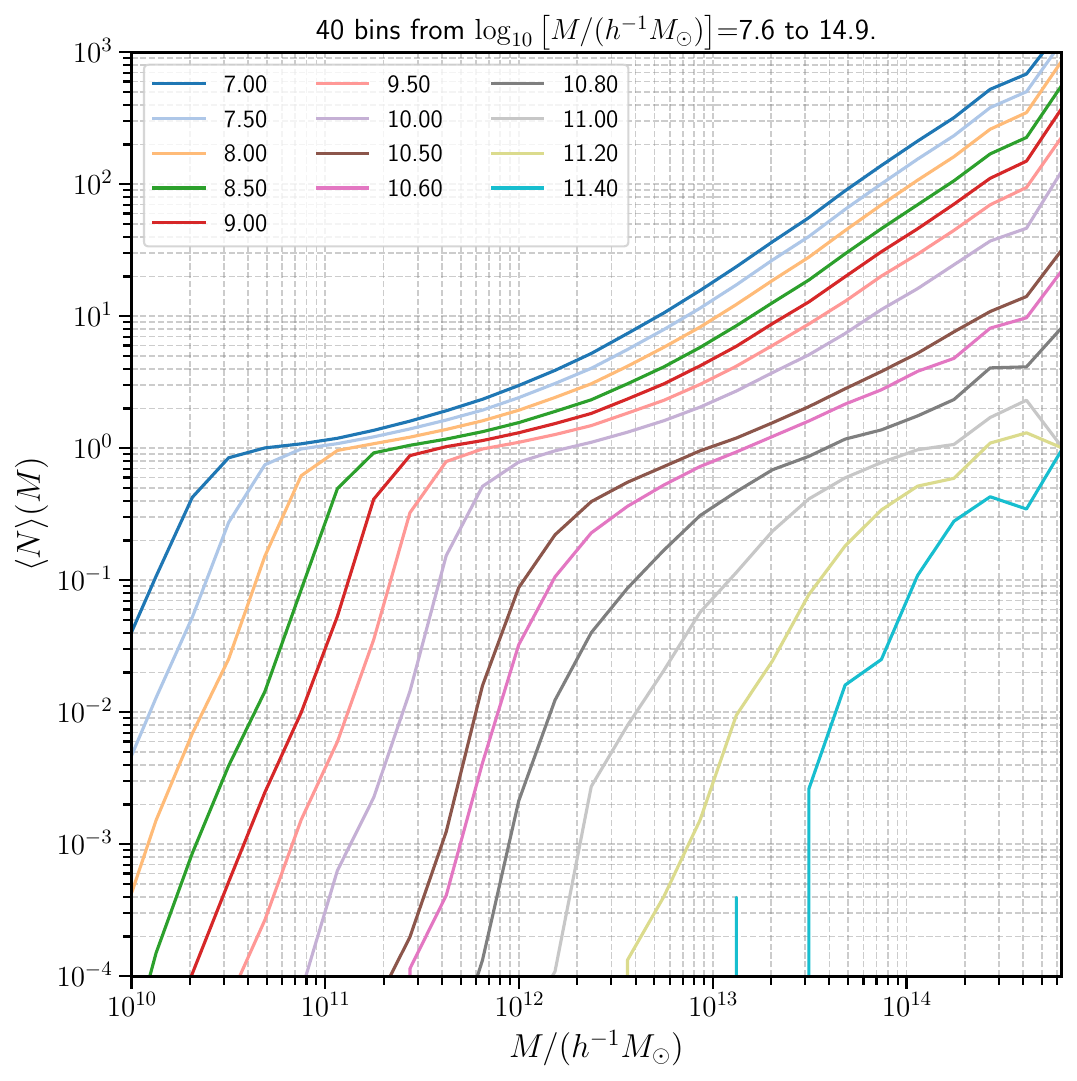}
        \caption{$\zt \in z_1$ bin}
        \label{fig:UTHODs_z1}
    \end{subfigure}
    \begin{subfigure}[b]{0.49\textwidth}
        \includegraphics[width=\columnwidth]{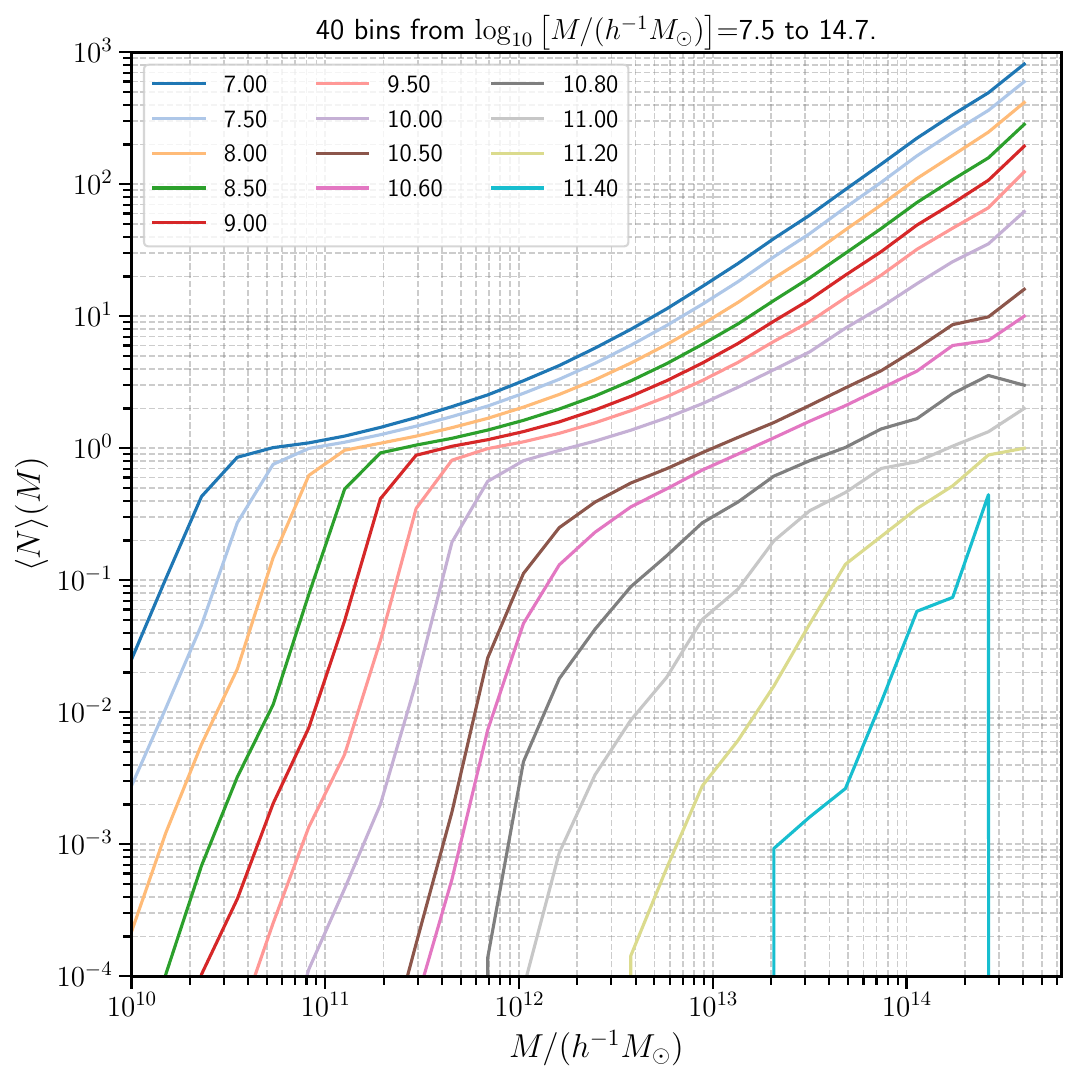}
        \caption{$\zt \in z_2$ bin}
        \label{fig:UTHODs_z2}
    \end{subfigure}
    \caption{Halo occupation distributions constructed from the UM-mocks in the two redshift bins as a function of stellar mass threshold, $\logMstarlimit$.}
    \label{fig:UTHODs}
\end{figure*}

We show the HODs obtained from the UM galaxy catalogue as a function of stellar mass thresholds, which were employed in our analysis in the two redshift bins in Fig.~\ref{fig:UTHODs}.   

\subsection{Model bias: incompatibility of UM galaxy catalogue and the standard HOD model} \label{app:modelbias}
\begin{figure*}
    \centering
    \includegraphics[width=\textwidth]{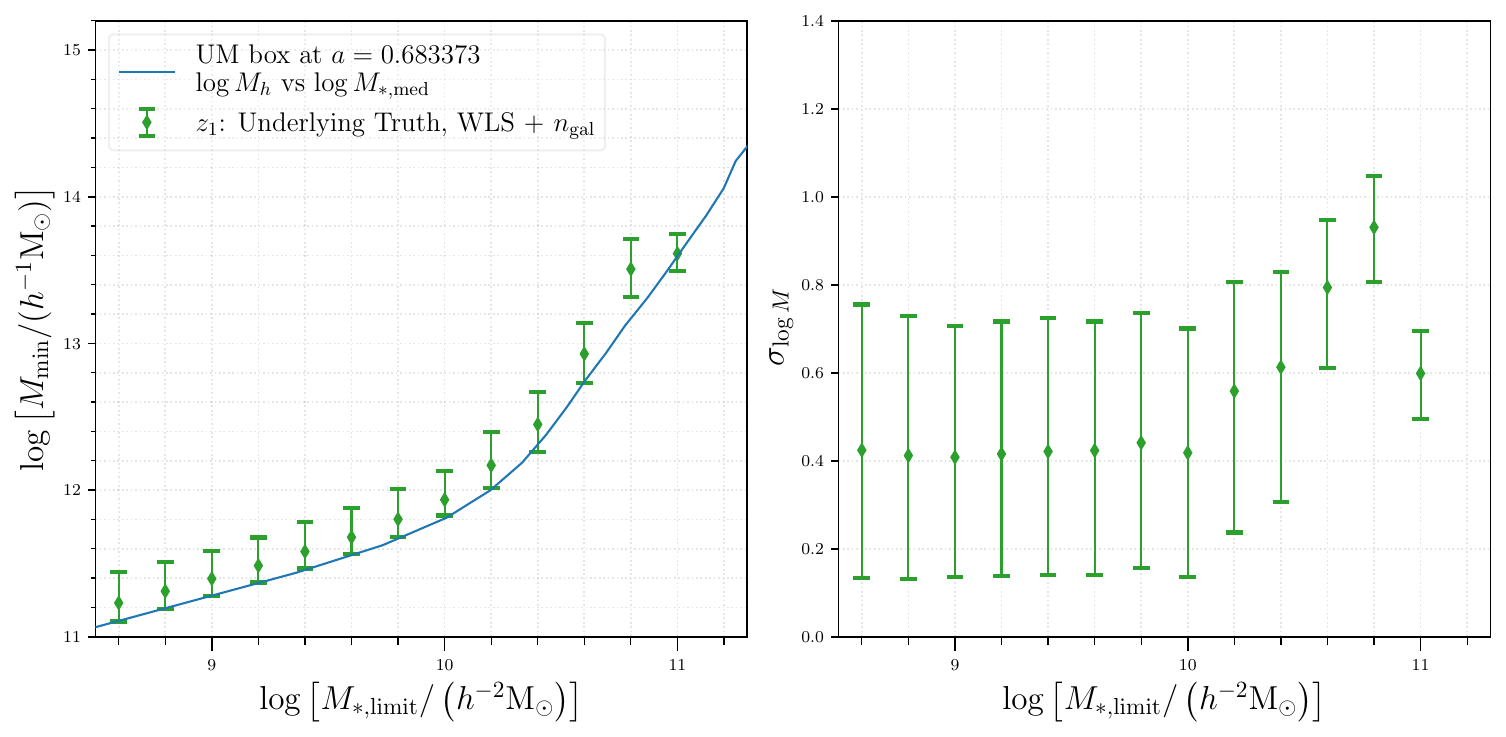}
    \caption{$z_1$: Central HOD parameters in UM-mock samples}
    \label{fig:z1_centralHOD_UT}
\end{figure*}

\begin{figure*}
    \centering
    \includegraphics[width=\textwidth]{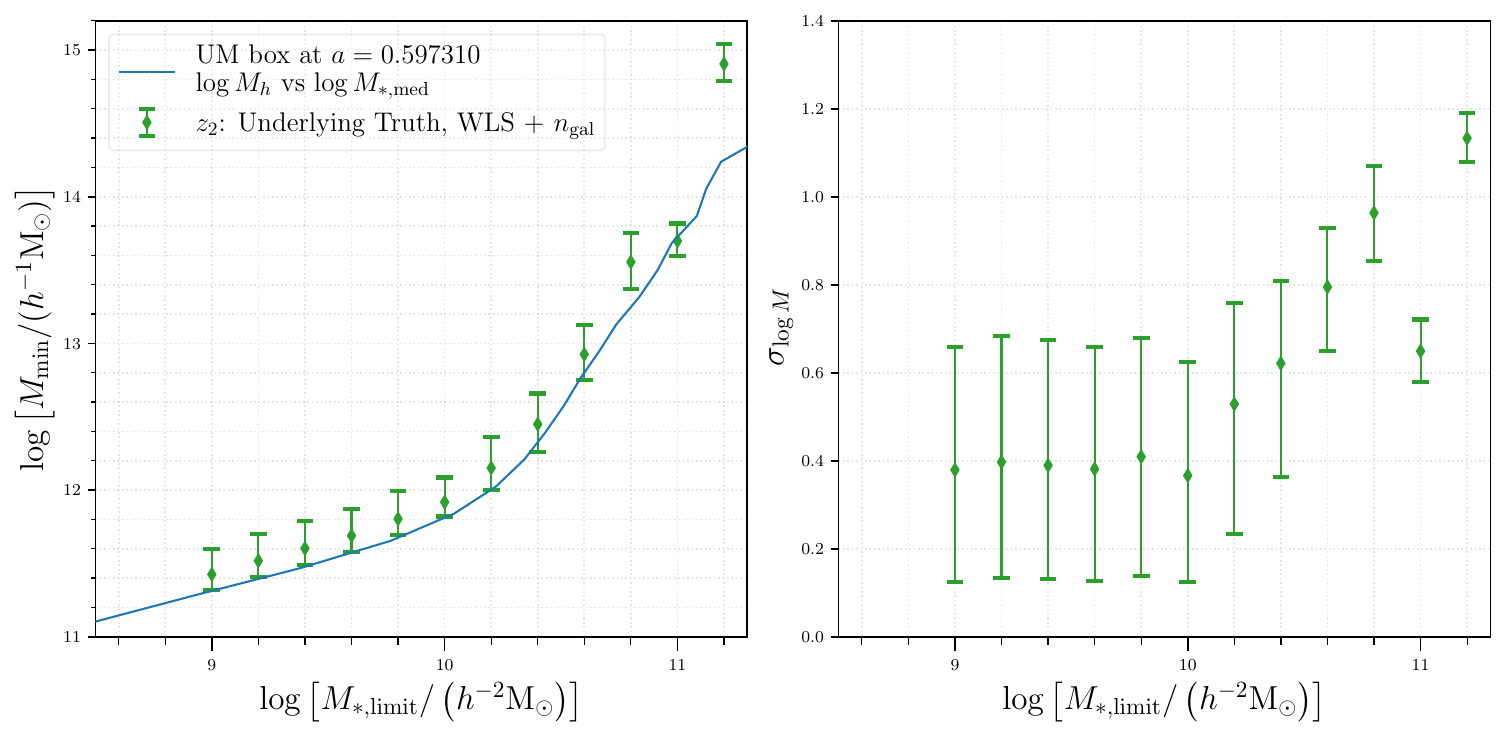}
    \caption{$z_2$: Central HOD parameters in UM-mock samples}
    \label{fig:z2_centralHOD_UT}
\end{figure*}

In this analysis, we have employed a widely used dark-matter only HOD model to fit the WLS obtained from the mocks of the UM galaxy catalogue. However, the form of central HOD model does not show enough freedom to fully capture the shape of the HOD measured directly from the mocks which most probably causes a bias in the SMHM relation of $\logMmin$ versus $\logMstarlimit$ which is seen to be true for both the redshift bins of our analysis in Figs.~\ref{fig:z1_centralHOD_UT} and \ref{fig:z2_centralHOD_UT} respectively. In the left-hand panel of each figure, the solid line represents the median stellar mass on the x-axis at various halo masses on the y-axis obtained using the UM-mocks, which is compared to the points and error bars, showing the median and the $1\sigma$ credible intervals for the $\logMmin$ parameter as inferred from our joint analysis of the UM-mocks. We find that the HOD modelling inferences of SMHM relation in the plane of $\logMmin$ parameter can be systematically biased upwards by as large as $1\sigma$ compared to the SMHM relation quoted in B19. 

We also show the corresponding constraint on the other central HOD parameter $\sigmalogM$ in the right-hand panel, as the lensing signal modelling is sensitive to a combination of two central HOD parameters.     

\bsp
\label{lastpage}
\end{document}